\documentclass[12pt]{article}
\usepackage{url}
\input{epsf}

\font\bm=cmmib10 at 10pt
\font\bms=cmmib10 at 7pt \textfont9=\bm \scriptfont9=\bms
\usepackage{amsfonts}
\usepackage{multirow}
\mathchardef\balpha= "790B
\mathchardef\bbeta= "790C
\mathchardef\bTheta= "7902
\mathchardef\bzeta= "7910
\mathchardef\bOmega= "790A
\mathchardef\bGamma= "7900
\mathchardef\bDelta= "7901
\mathchardef\bPhi= "7908
\mathchardef\bphi= "791E
\mathchardef\bomega= "7921
\mathchardef\bxi= "7918
\mathchardef\bet= "7911
\mathchardef\brho= "791A
\mathchardef\btau= "791C
\mathchardef\bmu= "7916
\mathchardef\bvarpi= "7924

\def \lvec{(\kern-.26em(}
\def\pmb#1{\setbox0=\hbox{#1}%
\def \lvec{(\kern-.26em(}
\kern-.025em\copy0\kern-\wd0
\kern.05em\copy0\kern-\wd0
\kern-.025em\raise.0433em\box0 }
\mathchardef\btheta= "7912
\usepackage{amsmath}
\usepackage{authblk}

\usepackage{graphicx}
\usepackage{dcolumn}

\begin{document}

\title{Nitrous Oxide and Climate}

\author[1]{ C. A. de Lange}
\author[2]{ J. D. Ferguson}
\author[3] {W. Happer}
\author[4]{ W. A. van Wijngaarden}
\affil[1]{Atomic, Molecular and Laser Physics, Vrije Universiteit, De Boelelaan 1081, 1081 HV Amsterdam, The Netherlands }
\affil[2]{University of Pennsylvania School of Veterinary Medicine, USA}
\affil[3]{Department of Physics, Princeton University, USA}
\affil[4]{Department of Physics and Astronomy, York University, Canada}
\renewcommand\Affilfont{\itshape\small}
\date{\today}
\date{\today}
\maketitle

\begin{abstract}
\noindent Higher concentrations of
atmospheric nitrous oxide  (N$_2$O) are expected to slightly warm Earth's surface because of increases in {\it radiative forcing}. Radiative forcing is the difference in the net upward thermal radiation flux from the Earth through a transparent atmosphere and radiation through an otherwise identical atmosphere with greenhouse gases. Radiative forcing, normally measured in W m$^{-2}$, depends on latitude, longitude and altitude, but it is often quoted for the tropopause, about 11 km of altitude  for temperate latitudes, or for the top of the atmosphere at around 90 km. For current concentrations of greenhouse gases, the radiative forcing per added N$_2$O molecule is about 230 times larger than the forcing per added carbon dioxide (CO$_2$) molecule. This is due to the heavy saturation of the absorption band of the relatively abundant greenhouse gas, CO$_2$, compared to the much smaller saturation of the absorption bands of the trace greenhouse gas N$_2$O.  But the rate of increase of CO$_2$ molecules, about 2.5 ppm/year (ppm = part per million by mole), is about 3000 times larger than the rate of increase of N$_2$O molecules, which has held steady at around 0.00085 ppm/year since the year 1985. So, the contribution of nitrous oxide to the annual increase in forcing is 230/3000 or about 1/13 that of CO$_2$.   If the main greenhouse gases, CO$_2$, CH$_4$ and N$_2$O have contributed  about 0.1 C/decade  of the warming observed over the past few decades, this would correspond to about $0.00064$ K per year or 0.064 K per century of warming from N$_2$O.   Proposals to place harsh restrictions on nitrous oxide emissions because of warming fears are not justified by these facts. Restrictions would cause serious harm; for example, by jeopardizing world food supplies.
\end{abstract}
\newpage
\tableofcontents

\newpage
\section{Introduction}
This is a sequel to an earlier paper from the CO$_2$ Coalition that discussed the small effects of increasing concentrations of atmospheric methane on Earth's climate, {\it Methane and Climate}\,\,\cite{methane}. The discussion below is focused on nitrous oxide.  There have been recent proposals to put harsh restrictions on human activities that   release nitrous oxide, most importantly, farming, dairying and ranching.

 Basic radiation-transfer science that is outlined in this paper gives no support to the assertion that greenhouse gases like nitrous oxide, N$_2$O, methane, CH$_4$, or carbon dioxide, CO$_2$,   are contributing to a climate crisis.  In fact, increasing concentrations of CO$_2$ have already benefitted the world by substantially increasing the yields of agriculture and forestry. For example, in a recent paper\,\cite{Taylor} Taylor and Shlenker state:
\begin{quote}
We find consistently high fertilization effects: a 1 ppm increase in CO$_2$ equates to a 0.5\%, 0.6\%, and 0.8\% yield increase for corn, soybeans, and wheat, respectively. Viewed retrospectively, 10\%, 30\%, and 40\% of each crop's yield improvements since 1940 are attributable to rising CO$_2$.
\end{quote}

Policies to address this non-existent crisis will cause enormous harm because of the vital role of nitrogen in agriculture.  The collapse of rice yields in Sri Lanka because of recent restrictions on nitrogen fertilizer should be a sobering warning\,\cite{SriLanka}. Much greater damage  will be done  in the future unless more rational policies are adopted.

We begin with a review of the limited degree to which nitrous oxide and other greenhouse gases can influence Earth's climate.  This is followed by a discussion of nitrogen's vital role in agriculture. This is a fairly technical paper, written primarily for scientists and  engineers; but we hope it will also be useful to non-technical readers and policy makers.

\begin{figure}[h]\centering
\includegraphics[height=80 mm,width=1\columnwidth]{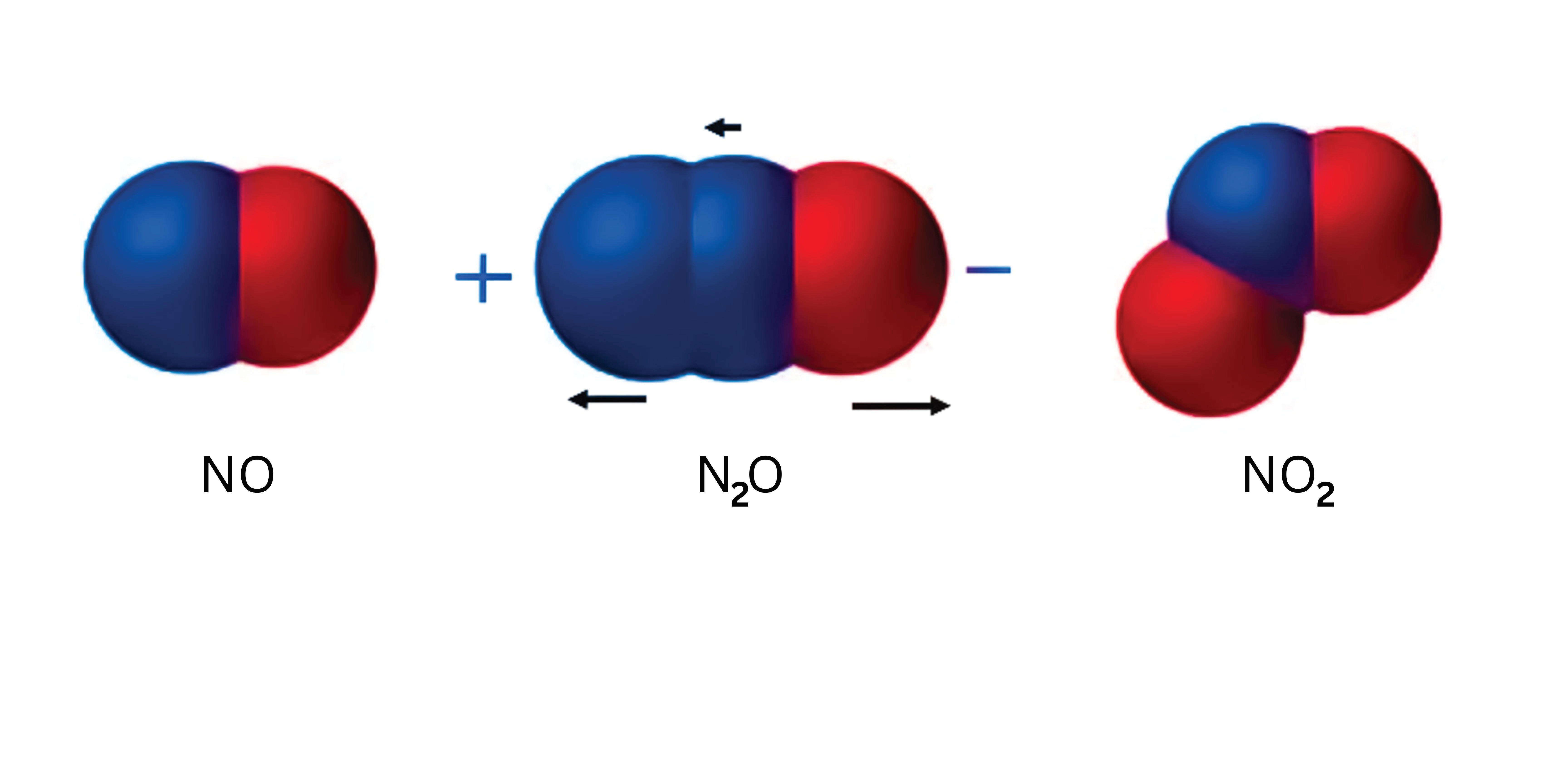}
\caption{\it The three most common oxides of nitrogen.  Only nitrous oxide (N$_2$O) is of concern as a greenhouse gas. About 60\% of the greenhouse forcing of N$_2$O comes from the linear stretch mode,  where the directions of atomic motion are indicated by the arrows, and where the vibrational frequency is $\nu_i = 1285$ cm$^{-1}$.  As indicated by the $+$ and $-$ signs, the N$_2$O molecule has a small permanent electric dipole moment that points from the negative O end (red) to the positive N end (blue). \label{N2O-molecule}}
\end{figure}

\begin{table}
\begin{center}
\begin{tabular}{||c c c c c c c ||}
 \hline
 Molecule & $\nu_i$ & $d_i$&$ n_i$& $P_i$ & $M_i$  & $\Gamma_{i}$\\
 &(cm$^{-1}$) &&&(10$^{-21}$ W)&(D)&(s$^{-1}$)\\ [0.5ex]
 \hline\hline
 H$_2$O&1595&1&$4.75\times 10^{-4}$&0.330&0.186&21.9\\
 &3652&1&$2.47\times 10^{-8}$&$6.40\times 10^{-5}$ & 0.068 & 35.7\\
 &3756&1&$1.50\times 10^{-8}$&$5.81\times 10^{-5}$&0.077&49.2\\
 \hline
 CO$_2$ & 667&2&$8.50\times 10^{-2}$& 1.73 & 0.181& 1.53\\
 &1388&1& $1.18\times 10^{-3}$ & 0 & 0 & 0 \\  
 & 2349&1& $1.18\times 10^{-5}$& 0.253 & 0.457& 424\\
 \hline
N$_2$O &588&2& $1.26\times 10^{-1}$& 0.224 & 0.069 &0.152  \\
&1285&1& $ 1.87 \times 10^{-3}$& 0.673 & 0.206 &14.1  \\
&2224& 1& $2.06\times 10^{-5}$&0.221 & 0.375 &243  \\
 \hline
 CH$_4$ &1311&3& $5.58\times 10^{-3}$&0.332  & 0.080 & 2.28 \\ 
 &1533&2& $1.28\times 10^{-3}$ & 0 & 0 & 0 \\  
  &2916&1& $8.38\times 10^{-7}$ & 0 & 0 & 0 \\ 
&3019&3& $1.53\times 10^{-6}$&$2.52\times 10^{-3}$ & 0.080 & 27.4 \\
\hline
\end{tabular}
\end{center}
\caption{\it Naturally occurring greenhouse gas molecules are listed in the first column. The vibrational mode frequencies $\nu_i$ are listed in the second column. The mode degeneracies $d_i$ are listed in the third column.  The mean numbers of vibrational quanta per molecule are listed in the fourth column for a temperature $T = 300$ K . At the  same temperature, the thermal powers $P_i$ radiated by molecules  at the frequencies of the $i$th mode are listed in the fifth column. The transition dipole moments $M_i$ for the modes are listed in the sixth column. The last column lists the spontaneous decay rates $\Gamma_i $ for molecules with one excitation quantum of the $i$th vibrational mode. The values of $\Gamma_i$ were taken from Table 5 of reference\,\cite{WH2}. There, one can also find details on how to calculate $\Gamma_i$ from the hundreds of thousands of line intensities and frequencies in the HITRAN data base\,\cite{HITRAN}.
\label{table1}}
\end{table}

\section{Radiative Properties of Greenhouse Gas Molecules}

Most of dry air, 99.96\% by volume, consists of the non-greenhouse gases nitrogen, oxygen and argon\,\cite{atmosphere}. For moist air, the concentration of water vapor, by far the most abundant greenhouse gas, is extremely variable, but it typically makes up several percent. The second most abundant greenhouse gas, carbon dioxide (CO$_2$) makes up about $0.04$\% of dry air, methane (CH$_4$) is $0.00019$\% and nitrous oxide (N$_2$O)  is only $0.000034$\%.  These seem negligibly small, but one must remember that greenhouse gases resemble food dyes. A little goes a long way. Tiny concentrations, most especially of water vapor and carbon dioxide, can greatly increase the opacity of air to thermal radiation that carries solar heat back to space. In the same way,  a few drops of green dye can color a much larger volume of nearly clear lager beer bright green on St. Patrick's Day.  But quantitative considerations  outlined below show that  future increases of nitrous oxide, N$_2$O, or methane, CH$_4$,  will have a negligible effect on climate.

Fig. \ref{N2O-molecule} shows the three  oxides of nitrogen commonly found in the air. Only
nitrous oxide (N$_2$O) has a significant greenhouse effect. Nitric oxide (NO), and nitrogen dioxide (NO$_2$) are also released by human activities, notably by high-temperature combustion of fossil fuels in air. At high concentrations NO, and especially NO$_2$, which readily reacts with water to form nitric acid, can have harmful effects on human health. But the greenhouse effects of NO and NO$_2$ are completely negligible.

Key radiative properties of N$_2$O and the other important naturally occurring greenhouse gases, methane (CH$_4$) carbon dioxide (CO$_2$) and most importantly, water vapor (H$_2$O), are summarized in Table \ref{table1}.
Ozone (O$_3$) is also an important greenhouse gas, but we have not included it in the table because most ozone is located in the stratosphere, and changes of its concentration there have little direct effect on surface temperatures.

The second column of Table \ref{table1} lists the normal-mode  vibration frequencies $\nu_i$.  For thermal infrared radiation, these are traditionally quoted in waves per centimeter (cm$^{-1}$).  The characteristic ways that the atoms of a molecule can vibrate about the center of mass are called normal modes. For simple diatomic molecules like nitrogen (N$_2$) or oxygen (O$_2$) the vibrating atoms simply stretch and compress the chemical bond between them. But for molecules with three or more atoms, there are more complicated vibrational modes, each with its own  vibrational frequency, $\nu_i$, and spatial pattern.  Instructive animations of the vibrational modes of N$_2$O, CH$_4$, CO$_2$ and H$_2$O,  can be found at the links of reference \,\cite{modes}. 

The spatial degeneracies, $d_i$ of the modes,  are shown in the third column of Table \ref{table1}.  Degeneracies are the number of ways a molecule  can vibrate at the same frequency but in different directions with respect to some reference orientation.  For example, the spatial degeneracy of the bending modes of CO$_2$ or N$_2$O are $d_i = 2$. If the reference orientation of the  linear molecules is vertical,  there can be bending vibrations, at the same frequency $\nu_i$, from left to right or forward and backward.  The tetrahedral molecule methane (CH$_4$) has some normal modes that are three-fold degenerate, with $d_i=3$.

 The free vibration and rotation of a molecule in the atmosphere is interrupted from time to time by collisions with other molecules and, much less frequently, by emission or absorption of photons.  Relatively large amounts of vibrational, rotational and translational energy are exchanged in each collision. The rapid collisions will make the probability of finding a molecule in a quantized state of energy $E_i$ proportional to the Boltzmann factor $e^{-E_i/kT}$, where $k$ is Boltzmann's constant and where the surrounding air has the local absolute temperature $T$.  Since the characteristic thermal energy $kT$ is much smaller than the quantized excitation energy $hc\nu_i$ of a mode with frequency $\nu_i$ (here $h$ is Planck's constant and $c$ is the speed of light), the average number of vibrational quanta $n_i$  is very small and is given (very nearly) by the Arrhenius factor,
\begin{equation}	
n_i=d_ie^{-hc\nu_i/kT}
\label{mol2}
\end{equation}
The mean numbers of excitation quanta $n_i$ listed in the  fourth column of Table \ref{table1} for a temperature of $T=300 $ K are within a few percent of the simple approximation (\ref{mol2}).

The electric fields of thermal radiation do work on moving charges within a molecule. If the work is positive, the molecule absorbs radiation and increases its energy at the expense of energy removed from the radiation.
If the work is negative, the molecule loses energy, which is emitted as additional radiation. The nuclei and electrons of the molecule produce a complicated charge density $\rho({\bf r}, t)$ at  locations ${\bf r}$ and time $t$. The positive charge is localized within the tiny atomic nuclei. The negative charges from electrons are much more delocalized. The chemical bonds of molecules, which are formed by valence electrons, are negatively charged.

The wavelengths of thermal infrared radiation are much longer than the sizes of molecules.  For example, the distance between the nucleus of the N atom on the left of the N$_2$O molecule of Fig. \ref{N2O-molecule}  and the nucleus of the O atom on the right is about 0.231 nm (nanometer, or $10^{-9}$ m). The wavelength of the main greenhouse mode of N$_2$O, with a vibrational frequency of $\nu_i = 1285$ cm$^{-1}$, is $\lambda_i = 1/\nu_i =$ 7,782 nm, some 34,000 times longer than the length of the molecule.  Under these conditions, the interaction of the molecules with radiation is almost completely controlled by the electric dipole moment, ${\bf M}(t)$ or the first moment of the charge density, which at time $t$ is given by
\begin{equation}	
{\bf M}(t)=\int{\bf r}\, \rho({\bf r},t)dV. 
\label{mol4}
\end{equation}
Here, $\rho({\bf r},t) dV = dq$ is the charge located in the infinitesimal volume $dV$ centered on the location ${\bf r}$ at time $t$.
Radiation generated by oscillations of the electric quadrupole, octupole, and higher moments of the molecule is many orders of magnitude  less powerful than that of the electric dipole moment ${\bf M}(t)$ of (\ref{mol4}) and can be ignored. Radiation from the oscillating magnetic moments of the molecule can also be ignored.

For electric-dipole radiation, the radiative power $P$ that a vibrating and rotating molecule emits can be calculated with Larmor's classic formula  \,\cite{Larmor} 
\begin{equation}	
P=\frac{2|{\bf \ddot M}|^2}{3c^3}
\label{mol6}
\end{equation}
Here ${\bf \ddot M}=d^2{\bf M}/dt^2 $, the second derivative with respect to time, is the acceleration of the dipole moment ${\bf M}$. Eq. (\ref{mol6}) is for cgs units.
Large, collision-induced changes of the molecular quantum states will cause the radiated power (\ref{mol6}) for an individual atmospheric molecule to fluctuate chaotically. But the  mean value of (\ref{mol6}), averaged over a time long enough to include many collisions, will have a well-defined average value $\langle P\rangle = P_1+P_2+P_3+\cdots$, where $P_1$ is the power radiated near the frequency $\nu_1$ of the first vibrational mode, $P_2$ is the power radiated near the frequency $\nu_2$ of the second, etc.  The mean radiated powers depend on the temperature $T$, and on the vibrational frequency $\nu_i$. The powers $P_i$ are displayed in the fifth column of Table \ref{table1}.  These are very small powers, of order $10^{-21}$ watts per molecule or less.  For comparison, a mobile phone radiates a few watts of power, and is not supposed to exceed 3 watts.

From inspection of Table \ref{table1},  we see that three of the powers $P_i$ in the fifth column are zero.  These correspond to modes for which the molecular charge density $\rho({\bf r},t)$ is so symmetric that vibrations and rotations do not produce corresponding vibrations of the dipole moment and, therefore, generate no radiation.
 For Table \ref{table1} these especially symmetric vibrational modes are the symmetric stretch mode of the CO$_2$ molecule, which vibrates with a frequency of 1388 cm$^{-1}$, and the modes of methane, CH$_4$ with frequencies $\nu_i$ of 1533 cm$^{-1}$ and 2916 cm$^{-1}$.

Molecules of the atmosphere's two most abundant gases, nitrogen (N$_2$) and oxygen (O$_2$) have vanishing dipole moments, ${\bf M}(t)=0$, because of their high symmetry and they therefore absorb or emit negligible amounts of thermal radiation. O$_2$ does emit and absorb millimeter-wave thermal radiation because of spin-flip transitions. Satellite observations of this radiation are used
for monitoring  the temperatures of Earth's atmosphere\,\cite{MSU}. But the heat transported by the millimeter waves is negligible.
 
The amplitude of a vibrating dipole moment of a molecule with $n_i$ quanta of vibrational excitation can be written as 
\begin{equation}	
M(t)=\sqrt{2n_i}M_i\cos\omega_i t.
\label{mol8}
\end{equation}
The angular frequency $\omega_i$ is related to the spatial frequency $\nu_i$ by
\begin{equation}	
\omega_i =2\pi\nu_i c.
\label{mol10}
\end{equation}
In (\ref{mol8}), $M_i\ge 0$ is the {\it transition electric dipole moment} of the molecule, the root-mean-square value  of the oscillating moment.  The number of vibrational quanta $n_i$ was given by the Boltzmann factor (\ref{mol2}).  Substituting (\ref{mol8}) into (\ref{mol6}) we find that the transition dipole moment is given by the (cgs) formula
\begin{equation}	
P_i=\frac{32\pi^4 \nu_i^4 c n_i M_i^2 }{3}.
\label{mol12}
\end{equation}
The average power $P_i$ radiated per molecule of (\ref{mol12}) increases very rapidly with temperature because the number of vibrational excitation quanta $n_i$ increases so rapidly, approximately as the Arrhenius factor of (\ref{mol2}).

Solving (\ref{mol12}) with the values of $P_i$, $n_i$ and $\nu_i$ gives the magnitudes  $|M_i|$ of the transition electric dipole moments listed in the sixth column of Table \ref{table1}.  Molecular dipole moments are traditionally measured in Debye units, where 1 D = $10^{-18}$ esu cm = $3.34\times 10^{-30}$ C m. 
The time needed for a molecule  to radiate away a quantum of vibrational energy turns out to be tens of milliseconds. This is orders of magnitude longer than the time between collisions with other molecules, around 1 nanosecond at sea-level pressures. Even for the extremely low air pressures of the upper stratosphere, the time between collisions, around 1 microsecond, is much shorter than the tens of milliseconds needed for a vibrating molecule to emit a photon.
Relatively large amounts of vibrational, rotational and translational energy are exchanged in each collision; and this leads to the Boltzman distribution (\ref{mol2}) of the molecules  in their quantized energy states.

The last column of Table \ref{table1} lists the spontaneous radiative decay rates $\Gamma_i$ of molecules with one quantum of vibrational excitation of frequency $\nu_i$. The inverse of this rate is the radiative lifetime, $\tau_i = 1/\Gamma_i$, the time needed for a molecule to radiate away $1-1/e = 63\,$\% of its vibrational energy if it is not interrupted by a collision with another molecule.
The  spontaneous decay rate $\Gamma_i$ is related to the other molecular parameters in the table by
\begin{equation}	
\Gamma_i=\frac{P_i}{h c\nu_i  n_i}=\frac{32\pi^4\nu_i^3 M_i^2}{3h}.
\label{mol14}
\end{equation}
\subsection{Permanent dipole moment}
\begin{table}
\begin{center}
\begin{tabular}{||c c c ||}
 \hline
 & $B$ & $M$\\
  Molecule&(cm$^{-1}$) &(D)\\ [0.5ex]
 \hline\hline
OH&18.91&1.67\\
 \hline
N$_2$O &0.419&0.161 \\
 \hline
\end{tabular}
\end{center}
\caption{\it Rotational constants $B$ of (\ref{mol22}) and permanent electric dipole moments $M$ of the diatomic hydroxyl molecule OH \,\cite{ROH} and the nitrous oxide molecule N$_2$O\,\cite{RN2O}.                
\label{table2}}
\end{table}

As indicated in Fig. \ref{N2O-molecule}, the N end of the linear N$_2$O molecule has a small positive charge and the O end has a small negative charge. The resulting permanent electric dipole moment, $M$, listed in Table \ref{table2},  is relatively small,
$M=0.161 $ D. 
 
Molecules with non-vanishing permanent electric dipole moments  can emit or absorb radiant energy at the expense of decreasing or increasing their rotational energy. They do not need to change their vibrational states.  The pure rotational transitions of the water molecule (H$_2$O) dominate the opacity of Earth's atmosphere for radiation frequencies less than about 500 cm$^{-1}$. But as we will outline below, the analogous pure rotational transitions of the N$_2$O molecule have a negligible effect on opacity.  
For more symmetric molecules like N$_2$, O$_2$, CO$_2$ or CH$_4$, the permanent electric dipole moments vanishes, $M=0$, and there is no pure rotational absorption or emission.

With respect to rotations, the water molecule (H$_2$O) is an asymmetric top, with three different principal moments of inertia. The details of rotational motion of asymmetric top molecules are relatively complicated, both for classical mechanics and quantum mechanics. So, for comparisons with the radiative properties of the linear N$_2$O molecule, we will consider the linear diatomic OH molecule, which has an electric dipole moment $1.67$ D that is similar\,\cite{Dipole} to that of H$_2$O, $M = 1.85$ D. The moment of inertia of OH is comparable to the three moments of inertia of H$_2$O. The  thermal radiation powers emitted and absorbed by pure rotational transitions of  OH and H$_2$O are about the same.

Both OH  and N$_2$O are  linear molecules, with moments of inertia $I$ for rotations about any axis normal to the symmetry axis of the molecule and through the center of mass. The spectroscopic ``rotation constant'' $B$ of the  molecules is related to the moment of inertia $I$ by
\begin{eqnarray}	
 B=\frac{\hbar^2}{2I h c}.
\label{mol22}
\end{eqnarray}
The rotational constants of N$_2$O and OH are listed in Table \ref{table2}.  For both molecules, the characteristic rotational energy $h c B$ is  much smaller than the thermal energy $kT$
\begin{equation}	
B\ll \frac{kT}{hc} \approx 200 \hbox{ cm}^{-1}.
\label{mol24}
\end{equation}
 Under these conditions both classical physics and quantum mechanics show that  the thermally averaged value of (\ref{mol6}) is 
\begin{eqnarray}	
P&=&\frac{16 k^2 T^2 M^2}{3I^2 c^3}.
\label{mol26}
\end{eqnarray}
Using the values of $B$ and $M$ from Table \ref{table1}, together with (\ref{mol22})  in (\ref{mol26}), we find that the pure rotational power $P(\hbox{N$_2$O})$ radiated by an N$_2$O molecule is more than five orders of magnitude smaller than the power $P(\hbox{OH})$, radiated by an OH molecule
\begin{equation}	
\frac{P(\hbox{N$_2$O})}{P(\hbox{OH})}=0.456 \times 10^{-5}.
\label{in40}
\end{equation}
So, although N$_2$O does have pure rotational emission and absorption, like H$_2$O, it is so small that it can be neglected; and we need only consider the vibration-rotation transitions, as is the case for CO$_2$ and CH$_4$. Water vapor is the only greenhouse gas for which the pure rotational band matters.

\section{Radiation Transfer in the Atmosphere}
The properties of individual greenhouse gases discussed in connection with Table 
\ref{table1} are not sufficient to understand Earth's greenhouse warming. There are so many greenhouse gas molecules that the radiation power  $P_i $ of (\ref{mol6}), emitted by one molecule, is very likely to be absorbed by another molecule, of the same or different chemical species, before the radiation can escape to space and cool the Earth.  The radiation to space comes from an {\it emission height}, $z^{\{e\}}(\nu)$, from which there is a high probability of escape to space because so few absorbing molecules remain overhead. The emission height, $z^{\{e\}}(\nu)$, has a complicated dependence on the radiation frequency $\nu$. For the weakly absorbed frequencies of the clear - sky atmospheric window between about 800  and 1200 cm$^{-1}$, the emission height can be taken to be zero $z^{\{e\}}(\nu)=0$ and the radiation to space comes directly from ground.  For the other extreme of frequencies that are strongly absorbed by greenhouse gases, the emission heights can be several km or greater. Nearly all the radiation to space comes from atmospheric molecules near the emission height. Most of the surface radiation is absorbed and does not reach outer space.

Radiation transfer in the cloud-free atmosphere of the Earth is controlled by only two quantities: (a) how the temperature $T=T(z)$ varies with the altitude $z$, shown in the left panel of Fig. \ref{GGNTa}, and (b) the altitude dependence of the molar concentrations, $C^{\{i\}}=C^{\{i\}}(z)$ of the $i$th type of molecule, shown on the right panel.
We will call $z$-dependent  quantities {\it altitude profiles}.  Although the altitude profiles of temperature and concentrations  vary with latitude and longitude,  the horizontal variation is normally small enough to neglect when calculating local radiative forcing.  The altitude profile of the temperature  is as important as the altitude profile of concentrations.  If the temperature were the same from the surface to the top of the atmosphere, there would be no radiative forcing, no matter how high the concentrations of greenhouse gases.
\subsection{Altitude profiles of molecular temperature}
\begin{figure}[h]\centering
\includegraphics[height=100mm,width=1\columnwidth]{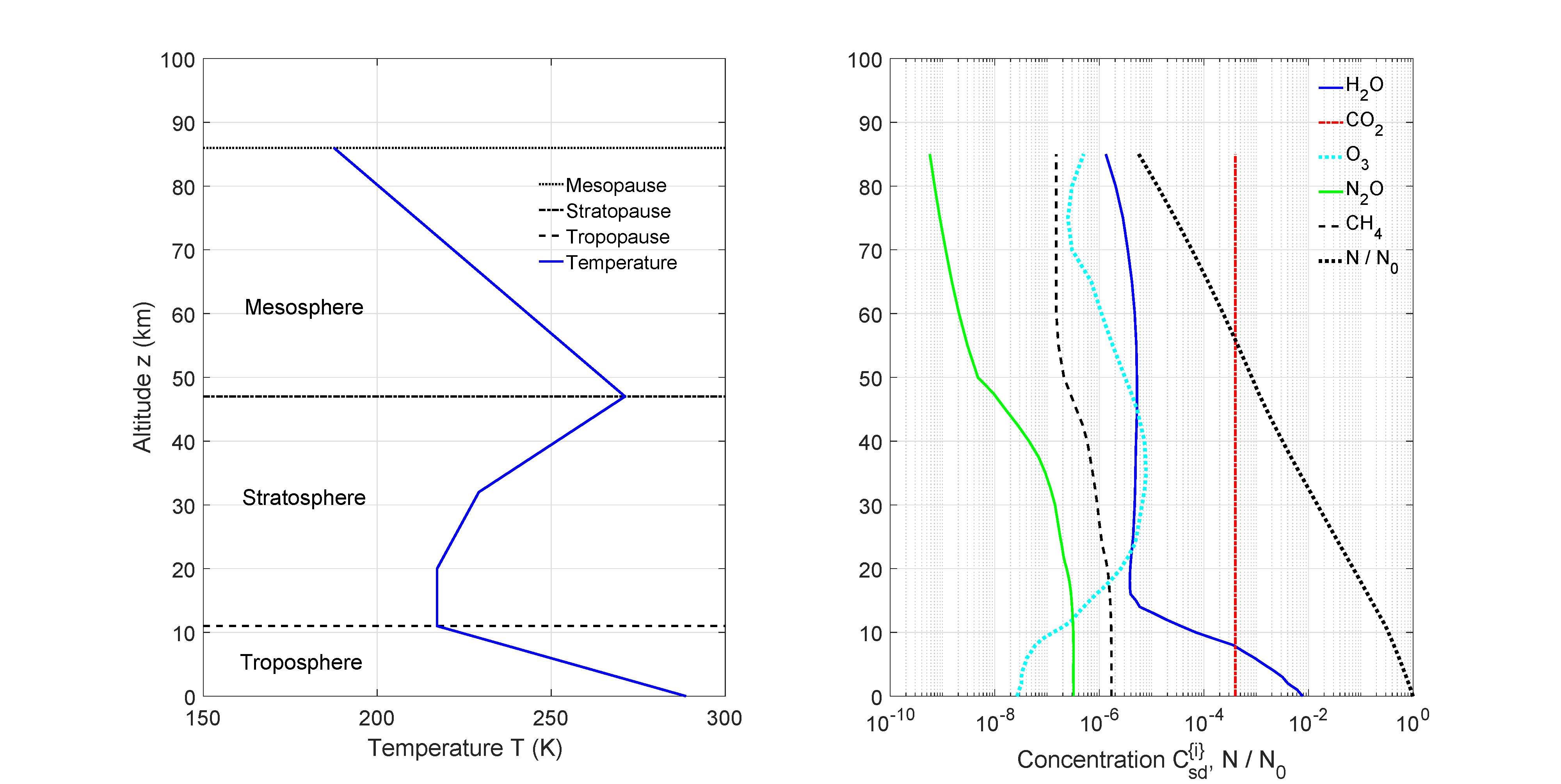}
\caption{\it {\bf Left.} A standard atmospheric temperature profile\,\cite{Temp}. The surface temperature is $T(0) = 288.7$ K.  {\bf Right.}  Concentration profiles\,\cite{Anderson}, $C^{\{i\}}_{\rm sd}=N^{\{i\}}_{\rm sd}/N$ for greenhouse molecules. The total number density of atmospheric molecules is $N=N(z)$. At sea level the concentrations are $7750$ ppm of H$_2$O, $1.8$ ppm of CH$_4$ and $0.32$ ppm of N$_2$O. The O$_3$ concentration peaks at $7.8$ ppm at an altitude of 35 km, and the CO$_2$ concentration was approximated  by $400$ ppm at all altitudes. The ratio $N/N_0$ of the total atmospheric number density, $N=N(z)$, to its surface value, $N_0=N(0)$, is also shown. \label{GGNTa}}
\end{figure}

Representative midlatitude altitude profiles of temperature \,\cite{Temp}, and concentrations of greenhouse gases\,\cite{Anderson}, are shown in Fig. \ref{GGNTa}. Altitude profiles of temperature directly measured by radiosondes in ascending balloons \,\cite{radiosonde} are always much more complicated than the profile in the left panel  of Fig. \ref{GGNTa}, which can be thought of as an ensemble average. As already implied by (\ref{mol2}), collision rates of molecules in the Earth's troposphere and stratosphere are sufficiently fast that for a given altitude $z$, a single local temperature $T=T(z)$ provides an excellent description of the distribution of molecules between translational, vibrational and rotational energy levels. 

On the left of Fig. \ref{GGNTa} we have indicated the molecular temperature of the three most important atmospheric layers for radiative heat transfer.   The lowest atmospheric layer is the troposphere, where parcels of air, warmed by contact with the solar-heated surface, float upward, much like  hot-air balloons. As they expand into the surrounding air, the parcels do work at the expense of internal thermal energy. This causes the parcels to cool with increasing altitude. There is very little heat flow in or out of the parcels during  ascent or descent, so expansions or compressions are very nearly adiabatic (or isentropic). If the parcels consisted of dry air, the cooling rate would be 9.8 C km$^{-1}$ the {\it dry adiabatic lapse rate}\,\cite{Lapse}. But rising air has usually picked up water vapor from the land or ocean, and the condensation of water vapor to droplets of liquid in clouds, or to ice crystallites, releases so much latent heat that the lapse rates can be much less than 9.8 C km$^{-1}$. A representative tropospheric lapse rate for midlatitudes is $-dT/dz = 6.5$ K km$^{-1}$ as shown in Fig. \ref{GGNTa}.
The tropospheric lapse rate is familiar to vacationers who leave hot areas near sea level for cool vacation homes at higher altitudes in the mountains.
On average, the temperature lapse rates are small enough to keep the troposphere buoyantly stable\,\cite{Buoyancy} so that higher-altitude cold air does not sink to replace lower-altitude warm air. Parcels of stable tropospheric air that are displaced in altitude will oscillate slowly up and down, with periods of a few minutes or longer. However, at any given time, large regions of the troposphere (particularly in the tropics) can become unstable to moist convection. Then small displacements of parcels can grow exponentially with time, rather than oscillating. 

Above the troposphere is the stratosphere, which extends from the tropopause to the stratopause, at a typical altitude of $47$ km, as shown in Fig. \ref{GGNTa}. Stratospheric air is much more stable to vertical displacements than tropospheric air, and negligible moist convection occurs there.  For midlatitudes, the temperature of the lower stratosphere is nearly constant, at about 220 K, but it increases at higher altitudes. At the stratopause the temperature reaches a peak value  not much less than the surface temperature. The stratospheric heating is due to the absorption of solar ultraviolet radiation by ozone molecules, O$_3$. The average solar flux at the top of the atmosphere is about 1350 Watts per square meter (W m$^{-2}$)\,\cite{Spectrum}. Approximately 9\% consists of ultraviolet light (with wavelengths shorter than $\lambda = 405$ nanometers (nm)) which can be absorbed in the upper atmosphere.

Above the stratosphere is the mesosphere, which extends from the stratopause to the mesopause at an altitude of about  $86$ km. With increasing altitudes above the mesopause, radiative cooling, mainly by CO$_2$ and O$_3$, becomes increasingly more important compared to heating by solar ultraviolet radiation. This causes the temperature to decrease with increasing altitude in the mesosphere.

Above the mesopause, is the extremely low-pressure thermosphere, where convective mixing processes are negligible.  Temperatures increase rapidly with altitude in the thermosphere, to as high as 1000 K, due to heating by extreme ultraviolet sunlight, the solar wind and atmospheric waves. Polyatomic molecules break up into individual atoms, and there is gravitational stratification, with lighter gases increasingly dominating at higher altitudes.

The vertical radiation  flux $ Z$, which is discussed below, can increase rapidly in the troposphere and less rapidly in the  stratosphere. The energy needed to increase the flux comes mostly from moist convection in the troposphere and mostly from absorption of ultraviolet sunlight in the stratosphere. There are still smaller increases of $Z$ in the mesosphere due to emission at the centers of the most intense lines of the greenhouse gases. Changes in $Z$ above the mesopause are small enough to be neglected, so we will often refer to the mesopause as ``the top of the atmosphere" (TOA), with respect to radiation transfer.
\subsection{Radiation temperature}
Radiation can have a temperature nearly equal to the molecular temperature if the mean-free paths of all photons are small compared to distances over which there is negligible variation of the molecular temperature.
The spectral intensity $\tilde I$ of radiation in full thermal equilibrium at a temperature $T$  is exactly equal to the Planck intensity \,\cite{Planck}
\begin{equation}	
\tilde B =\frac{2hc^2\nu^3}{e^{\nu c\, h/(kT)}-1}.
\label{ap2}
\end{equation}
The Planck spectral intensity $\tilde B$ depends on the spatial frequency $\nu$, measured in wavenumbers (cm$^{-1}$) and on the absolute temperature $T$ in Kelvin (K). The cgs units of $\tilde B$ are ergs per second, per square centimeter, per wavenumber, and per steradian of solid angle.  Eq. (\ref{ap2}) for radiation (photon) energy is closely related to the Boltzman distribution (\ref{mol2}) of molecules over their quantized energy states. Radiation of temperature $T$ must have the frequency spectrum (\ref{ap2}) and it must be isotropic, with equal intensities in all directions. 

Planck radiation in full thermal equilibrium is not commonly observed in Earth's atmosphere.  
The classic way to generate Planck radiation is with a {\it Hohlraum}, a cavity with opaque, isothermal walls and a tiny hole to allow observation of the radiation inside. 
So, Planck radiation is often called cavity radiation or blackbody radiation\,\cite{blackbody}.

The integral of the Planck intensity (\ref{ap2}) over all frequency increments $d\nu$,  and over all solid angle increments, $2\pi d\mu$ (radiation propagating at an angle $\theta$ to the zenith has the direction cosine $\mu=\cos\theta$)  gives the well-known Stefan Boltzmann flux, which was discovered several decades before the Planck spectrum (\ref{ap2})
\begin{equation}	
Z=\int _0^{\infty} \pi \tilde B d\nu =\sigma T^4 .
\label{ap4}
\end{equation}
Here, the Stefan-Boltzmann constant (for MKS units) is
\begin{equation}	
\sigma = 5.67 \times 10^{-8} \hbox{ W m$^{-2}$ K$^{-4}$}.
\label{ap6}
\end{equation}

Radiation temperatures are routinely measured with radiometers. These instruments detect the radiation flux $Z_r$ coming from some direction. The instrument is calibrated in a cyclic manner by interleaving views of the unknown radiation source with the flux $Z$ of a calibrating black body. The blackbody temperature $T$ is close to the expected radiation temperature.  For satellite-based radiometers\,\cite{MSU}, the calibration cycle normally includes an observation of dark space, which defines the zero-flux output of the detector. The radiometer gives an apparent, frequency-integrated radiation temperature
\begin{equation}	
T_r=\left(\frac{Z_r}{Z}\right)^{1/4}T.
\label{ap8}
\end{equation}
Some radiometers measure only environmental radiation and calibrating radiation with frequencies close to $\nu$. These  give a $\nu$-dependent apparent temperature, $T_r(\nu)$.

Thermal radiation in Earth's atmosphere is almost never in full thermal equilibrium. For cloud-free skies, the mean-free paths of photons can exceed the atmospheric thickness in the infrared atmospheric window from about 800  to 1300 cm$^{-1}$. For these frequencies, the very weak downwelling radiation from cold space will have an apparent temperature $T_r$ that is much smaller than the apparent temperature, $T_r$, of upwelling radiation from the relatively warm lower atmosphere and from Earth's surface. Because greenhouse gases absorb radiation in fairly narrow bands of frequencies, the frequency spectrum of the radiation in cloud-free skies differs drastically from that of the spectral Planck intensity (\ref{ap2}) that is required for radiation in full thermodynamic equilibrium.

The mean-free paths of thermal radiation photons inside thick clouds can be much shorter than the size of the cloud. There is also little frequency dependence to the absorption and emission coefficients of thermal radiation by  the water droplets or the  ice crystallites of the cloud. Cloud interiors are the only parts of the atmosphere where radiation is almost in thermal equilibrium, nearly isotropic and with a spectrum close to that of (\ref{ap2}).

Planck's spectral intensity (\ref{ap2}) is one of the most famous equations of physics. It finally swept aside the absurd prediction of pre-quantum physics that thermal radiation would have infinite intensity (the ultraviolet catastrophe), and it gave birth to quantum mechanics \,\cite{Planck, blackbody,ultraviolet}.  

\subsection{Altitude profiles of greenhouse gases}
As shown in Fig. \ref{GGNTa}, the most abundant greenhouse gas at the surface is water vapor.  However, the concentration of water  vapor drops by a factor of a thousand or more between the surface and the tropopause.  This is because of condensation of water vapor into clouds and  eventual removal by precipitation.

Carbon dioxide, CO$_2$, the most abundant greenhouse gas after water vapor,  is also the most uniformly mixed because of its immunity to further oxidation and resistance to photodissociation in the upper stratosphere.

Methane is  much less abundant than CO$_2$. Methane's concentration decreases somewhat in the stratosphere because of oxidation by  OH radicals and ozone, O$_3$. The oxidation of methane provides about 1/3 of the stratospheric water vapor shown in Fig.~\ref{GGNTa}, with most of the rest directly injected by upwelling in the tropics\,\cite{Fueglistaler}.

Nitrous oxide (N$_2$O),  the main topic of this discussion, is also much less abundant than CO$_2$, and its concentration decreases in the stratosphere because absorption of ultraviolet sunlight dissociates N$_2$O  to nitric oxide molecules (NO) and free atoms of O.

Ozone molecules (O$_3$) are produced from O$_2$ molecules by ultraviolet sunlight in the upper atmosphere, and this is the reason that O$_3$ concentrations peak in the stratosphere and are hundreds of times smaller in the troposphere, as shown in 
Fig. \ref{GGNTa}.

\begin{figure}[h]\centering
\includegraphics[height=88mm,width=1\columnwidth]{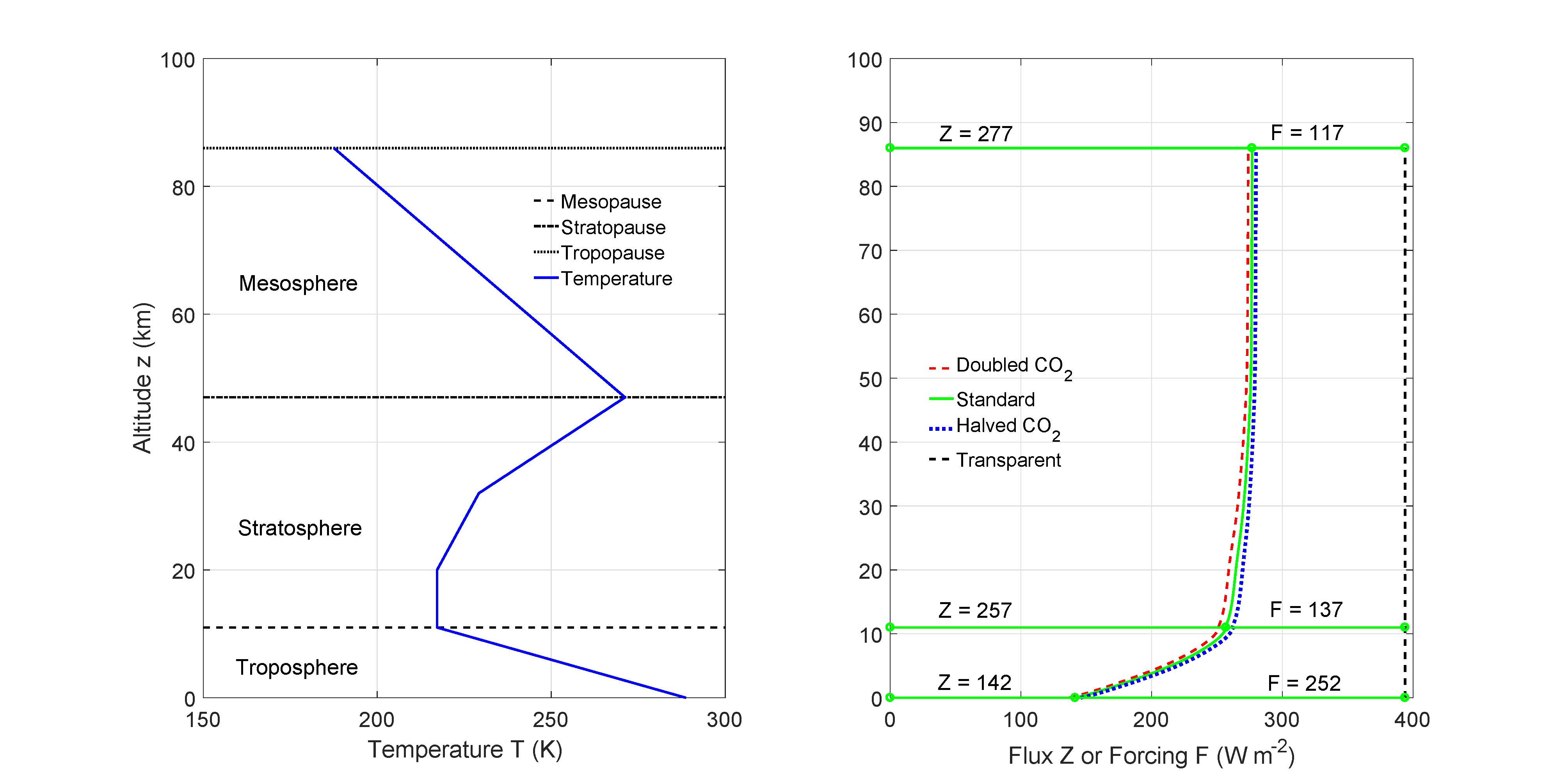}
\caption{\it {\bf Left:} The altitude dependence of temperature from Fig. \ref{GGNTa}. {\bf Right} The flux $Z$ increases with increasing altitude as a result of net upward energy radiation from the greenhouse gases H$_2$O,  O$_3$, N$_2$O, CH$_4$, and CO$_2$ in clear air with no clouds. The middle, green curve is the flux for current concentrations.  The forcings $F$ are the differences between the altitude-independent flux $Z_0=\sigma T_0^4$ through a transparent atmosphere with no greenhouse gases, for a surface temperature of $T_0 = 288.7$ K,  and the flux $Z$ for an atmosphere with the greenhouse gas concentrations of Fig. \ref{GGNTa}. Fluxes and forcings for halved and doubled concentrations of CO$_2$, but with the same concentrations of all other greenhouse gases, are shown as dotted blue and dashed red curves, which barely differ from the green curve, the flux for current concentrations.    We used doubled and halved  CO$_2$ rather than N$_2$O  for this illustration since the flux changes for doubling or halving N$_2$O  concentrations are about ten  times smaller than the corresponding fluxes for doubled and halved CO$_2$, and they would be too small to discern from the figure. From reference\,\cite{WH1}. \label{ZzCMa}}
\end{figure}

\subsection{Radiation intensities and fluxes}
The most detailed description of radiation in the atmosphere is given by the spectral intensity $\tilde I=\tilde I(z,\mu,\nu)$, which describes radiation of frequency $\nu$ at an altitude $z$, propagating in a direction that makes an angle $\theta$ to the vertical, with the direction cosine $\mu=\cos\theta$. For cloud-free skies there can be absorption and emission of thermal radiation by greenhouse gases, but there is negligible scattering. The propagation of the radiation is given by the Schwarzschild equation\,\cite{RadTrans}
\begin{equation}	
\mu\frac{\partial}{\partial z}\tilde I(z,\mu,\nu)=\alpha(z,\nu)\left[\tilde B(z,\nu)-\tilde I(z,\mu,\nu)\right].
\label{rif2}
\end{equation}
We assume that the molecular temperature $T(z)$ depends on altitude $z$. So, the Planck intensity $\tilde B(z,\nu)$ can be thought of as a function of altitude $z$ and spatial frequency $\nu$. The Planck intensity is equal for all directions and does not depend on the direction cosine $\mu$ of the radiation. The attenuation coefficient $\alpha(z,\nu)$ describes absorption and emission by greenhouse gases.  The Schwarzschild equation  of (\ref{rif2}) says that the actual intensity $\tilde I$ of radiation in the atmosphere is always trying to become equal to the local Planck intensity $\tilde B$, which would make the right side of the equation vanish,  and stop any further changes of $\tilde I$.  As explained in detail in reference\,\cite{WH1}, the Schwarzschild equation (\ref{rif2}) has a well-defined solution that is determined by the temperature profile $T(z)$ and the attenuation-rate profile 
$\alpha(z,\nu)$ of the atmosphere, together with emission of radiation by Earth's surface. It is worth stressing that the altitude dependence of temperature $T(z)$ is as important as altitude dependence of greenhouse gas concentrations which determines $\alpha(z,\nu)$.

For discussions of climate, the upward spectral flux 
\begin{equation}	
\tilde Z(z,\nu)=\int_{-1}^1 d\mu \,\mu \tilde I(z,\mu,\nu)
\label{rif4}
\end{equation}
is of more importance than the intensity $\tilde I$. The flux, $\tilde Z(z,\nu) d\nu$, measures the energy flow, in units of W m$^{-2}$ cm, for radiation of frequencies between $\nu$ and $\nu+d\nu$ at the altitude $z$ in the atmosphere. In (\ref{rif4}) the parts of the integrand with upward direction cosines, $\mu>0$ are called upwelling. They come from the Earth's surface emissions and from the emission of greenhouse gases at altitudes lower than the observation altitude $z$.  The parts of the integrand with downward direction cosines, $\mu<0$ are called downwelling. They come from the emission of greenhouse gases at altitudes higher than the observation altitude $z$.

\subsection{Forcings}
How  greenhouse gases affect radiative energy transfer through Earth's atmosphere, and ultimately Earth's climate, is quantitatively determined by the  {\it radiative forcing}, $F$, the difference between the flux $Z_0$  of thermal radiant energy from a black surface through a hypothetical, transparent atmosphere, and the flux $Z$ through an atmosphere with greenhouse gases, particulates and clouds, but with the same surface temperature $T_0$\,\cite{WH1}
\begin{equation}	
F=Z_0-Z.
\label{rff2}
\end{equation}
The flux from a black surface is given by the Stefan-Boltzmann formula
\begin{equation}	
Z_0=\sigma T_0^4.
\label{rff3}
\end{equation}
The forcing $F$ and the flux $Z$ are usually specified in units of W m$^{-2}$.  According to (\ref{rff2}), increments $\Delta F$ in forcing and increments $\Delta Z$ in net upward flux are equal and opposite 
\begin{equation}	
\Delta F=-\Delta Z.
\label{rff5}
\end{equation}
Forcing depends on latitude, longitude and on the altitude, $z$.

The radiative heating rate\,\cite{WH1},
\begin{equation}	
R=\frac{dF}{dz},
\label{rff6}
\end{equation}
is equal to the rate of change of the forcing with increasing altitude $z$.   Over most of the atmosphere, $R<0$, so thermal infrared radiation is a cooling mechanism that transfers internal energy of atmospheric molecules to space or to the Earth's surface. 

The definition (\ref{rff2}) of forcing is sometimes called the ``instantaneous forcing,'' since it  assumes that the concentration of the greenhouse gas of interest is instantaneously changed, but that all other atmospheric properties remain the same as before. Because the radiation itself travels at the speed of light even an ``instantaneous'' change in greenhouse gas concentrations requires a few tens of microseconds to reestablish radiative equilibrium. 

We consider the instantaneous forcing to be the least ambiguous metric for how changing greenhouse gases affect radiative transfer. But
the IPCC commonly uses a slightly different definition of radiative forcing (RF), where the stratosphere is allowed to cool to a new state of radiative equilibrium after  instantaneous addition of a greenhouse gas. In Section 8.1.1.1 of the 2018 IPCC report\,\cite{IPCC}, one finds a definition, identical to ours, of the instantaneous forcing:
\begin{quote}
Alternative definitions of RF have been developed, each with its own
advantages and limitations. The instantaneous RF refers to an instantaneous change in net (down minus up) radiative flux (shortwave plus
longwave; in W m$^{-2}$) due to an imposed change. This forcing is usually
defined in terms of flux changes at the top of the atmosphere (TOA)
or at the climatological tropopause, with the latter being a better indicator of the global mean surface temperature response in cases when
they differ.
\end{quote}

The right panel of Fig. \ref{ZzCMa} shows the altitude dependence of the net upward flux $Z$ and the forcing $F$ for the  greenhouse gas concentrations of Fig. \ref{GGNTa}. The temperature profile of Fig \ref{GGNTa} is reproduced in the left panel.  The altitude-independent flux, $\sigma T_0^4 = 394 $ W m$^{-2}$, from the surface with a temperature $T_0 = 288.7$ K, through a hypothetical transparent  atmosphere, is shown as the vertical dashed line in the panel on the right. The fluxes for current concentrations of CO$_2$ and for doubled or halved concentrations are shown as the continuous green line, the dashed red line  and the dotted blue line.

At current greenhouse gas concentrations the net upward flux at the surface, 142 W m$^{-2}$, is less than half the surface flux, $Z_0=\sigma T_0^4= 394\hbox{ W m}^{-2}$, also given by (\ref{sfm8}), for a transparent atmosphere. More than half of the upward thermal radiation from the surface is canceled by downwelling radiation from greenhouse gases above. At the tropopause altitude, 11 km in this example, the net flux has nearly doubled to $Z=257$ W m$^{-2}$. The 115 W m$^{-2}$ increase in flux from the surface to the tropopause is due to net upward radiation by greenhouse gases in the troposphere.  Most of the energy needed to supply the radiated power comes from convection of moist air. Direct absorption of sunlight in the troposphere makes a much smaller contribution.

From Fig. \ref{ZzCMa} we see that the flux $Z$ increases by another 20 W m$^{-2}$, from 257 W m$^{-2}$ to 277 W m$^{-2}$ between the tropopause and the top of the atmosphere.
The energy needed to supply the  20 W m$^{-2}$ increase in flux comes from the absorption of solar ultraviolet light by ozone, O$_3$ in the stratosphere and mesosphere. Convective heat transport above the tropopause is small enough to be neglected.

\begin{figure}[t]\centering
\includegraphics[height=100mm,width=1\columnwidth]{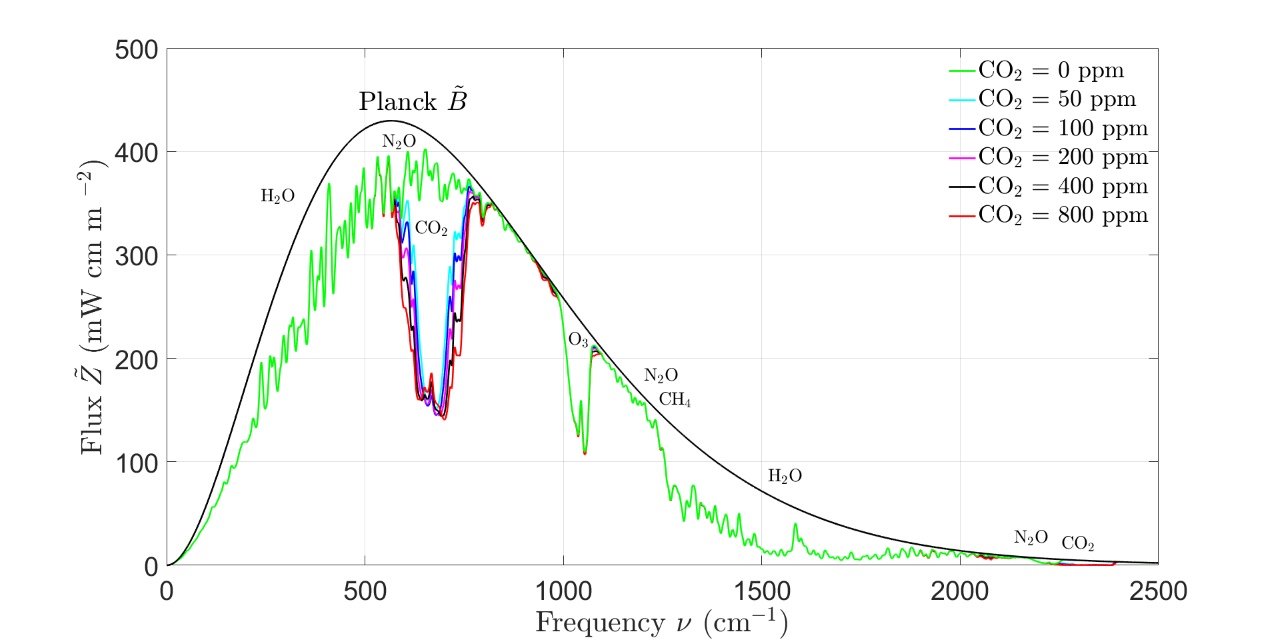}
\caption{\it Spectral radiation flux $\tilde Z$ at the top of the cloud-free atmosphere versus instantaneous changes of CO$_2$ concentrations from a reference value of 400 ppm, by factors of $2$ or $1/2$.   Because of ``saturation,'' every doubling (100\% increase) reduces the frequency-integrated flux to space by $\Delta Z = -3.0 \hbox{ W m}^{-2}$ and every halving (50\% decrease) increases the radiation to space by $\Delta Z = +3.0 \hbox{ W m}^{-2}$. The normal mode frequencies of Table \ref{table1}, where the greenhouse gases are most effective, are marked by the molecular formulas. Pure rotational transitions are responsible for the low-frequency absorption of water vapor, H$_2$O. The smooth  blue envelope curve is the flux to space,  $\pi \tilde B$, if there were no greenhouse gases.  The Planck intensity $\tilde B$ was given by 
 (\ref{ap2}) and the factor of $\pi$ comes from integrating the flux over upward solid angles.
 Computational details can be found in reference\,\cite{WH1}.  
\label{Corkforcingfigure}}
\end{figure}
\begin{figure}[t]\centering
\includegraphics[height=100mm,width=1\columnwidth]{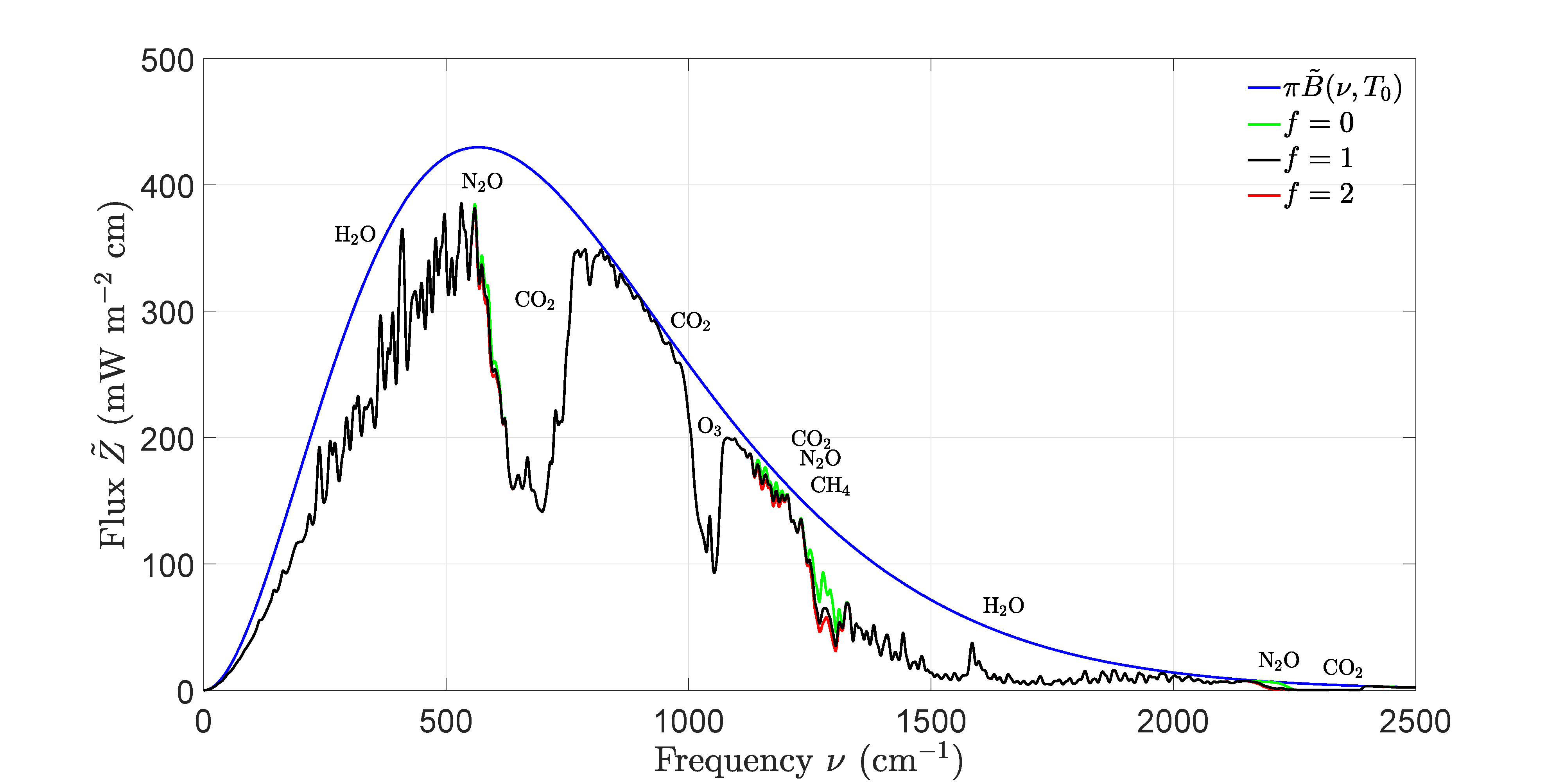}
\caption{\it The jagged black curve is the spectral flux $\tilde Z$ of (\ref{sfm2}) at the top of the cloud-free atmosphere for the conditions of Fig. \ref{GGNTa}. The jagged green curve shows the flux with all N$_2$O instantaneously removed,  but with no other changes to the conditions of Fig. \ref{GGNTa}.    The jagged red curve shows the  flux that results when the N$_2$O concentration is instantaneously doubled.  The barely perceptible differences between the green, black and red curves occur for frequencies close to N$_2$O's normal mode frequencies, that are listed in Table \ref{table1}. The contributions of the pure rotational band, at frequencies below about 500 cm$^{-1}$ are too small to perceive. The smooth  blue envelope curve is the flux to space,  $\pi \tilde B$, if there were no greenhouse gases.  The Planck intensity $\tilde B$ was given by 
 (\ref{ap2}) and the factor of $\pi$ comes from integrating the flux over upward solid angles.
Computational details can be found in reference\,\cite{WH1}. 
\label{N2O}}
\end{figure}

The fluxes, $Z$, and forcings, $F$,  of (\ref{rff2}) can be thought of as sums of infinitesimal contributions, $\tilde Z\, d\nu$ and $\tilde F\, d\nu$,  from {\it spectral fluxes}, $\tilde Z$, or {\it spectral forcings}, $\tilde F$,  carried by infrared radiation of spatial frequencies between $\nu$ and $\nu+d\nu$. 
That is,
\begin{equation}	
Z=\int_0^{\infty}\tilde Z\, d\nu = 277 \hbox{ W m}^{-2},
\label{sfm2}
\end{equation}
and
\begin{equation}	
F=\int_0^{\infty}\tilde F\, d\nu = 117 \hbox{ W m}^{-2}.
\label{sfm4}
\end{equation}
The integral (\ref{sfm2}) is the area under the jagged black curve of  Fig. \ref{N2O}. 
The spectral fluxes and forcings are related by a formula analogous to (\ref{rff2})
\begin{equation}	
\tilde F=\pi \tilde B_0 -\tilde Z.
\label{sfm6}
\end{equation}
Here  $\tilde B_0 =\tilde B(\nu,T_0)$, is the surface value of the spectral  Planck intensity (\ref{ap2}).

Analogous examples of how spectral fluxes $\tilde Z$ depend on greenhouse gases  are shown in Fig. \ref{Corkforcingfigure} for various concentrations of CO$_2$. To guard against an overly simplistic interpretation of this graph, the following should be realized. The flux at the top of the atmosphere comes from different altitudes at different frequencies.  Only in the infrared atmospheric window, from about 800 to 1200 cm$^{-1}$ is emission from Earth's surface directly observed, as long as no clouds get in the way. Stratospheric ozone obscures some of the atmospheric window at frequencies around 1050 cm$^{-1}$.

Fig.  \ref{Corkforcingfigure} is for cloud-free skies. Except in the atmospheric window, the infrared radiation observed by satellites is not surface radiation with various amounts of attenuation by greenhouse gases. Instead, it is newly created radiation that has been emitted at altitudes well above the surface, at the ``emission height."  The upwelling ground radiation has been completely absorbed. The radiation  in the frequency band centered on  667 cm$^{-1}$ is emitted by CO$_2$ molecules located in the nearly isothermal lower stratosphere, from altitudes between about 10 km and 20 km, where the temperature is about 220 K. 

Even if all the CO$_2$ were removed from the atmosphere, as shown by the jagged green line of Fig.  \ref{Corkforcingfigure},  no surface radiation would reach the top of the atmosphere in the band centered at 667 cm$^{-1}$ since there is substantial absorption by water vapor in this same band.  So, if all the CO$_2$ could be removed, but the atmospheric temperature profile and the concentrations of other greenhouse gases remained the same as for Fig. \ref{GGNTa}, one would observe emission of water vapor at a few km altitude, where it is warmer than the lower stratosphere, but colder than the surface. 

How doubling the current concentration of N$_2$O or removing it all  affects the  spectral flux at the top of the atmosphere is illustrated in  Fig. \ref{N2O}.  
The area under the black, jagged curve is $Z=277$ W m$^{-2}$ and is the frequency-integrated upward flux at the top of the atmosphere of Fig. \ref{ZzCMa}.  

The Stefan-Boltzman flux, $Z_0=\sigma T_0^4 = 394$ W m$^{-2}$ of (\ref{rff2}),  for a surface temperature of $T_0=288.7$ K,  is the frequency integral of the Planck spectral flux, $\pi \tilde B_0$, of (\ref{ap2})
\begin{equation}	
\int _0^{\infty} \pi \tilde B_0 d\nu =\sigma T_0^4 = 394 \hbox{ W m}^{-2}.
\label{sfm8}
\end{equation}
The integral (\ref{sfm8}) is the area in  Fig. \ref{N2O} beneath the smooth blue curve, the spectral flux for a transparent atmosphere.
\begin{figure}[h]\centering
\includegraphics[height=100mm,width=1\columnwidth]{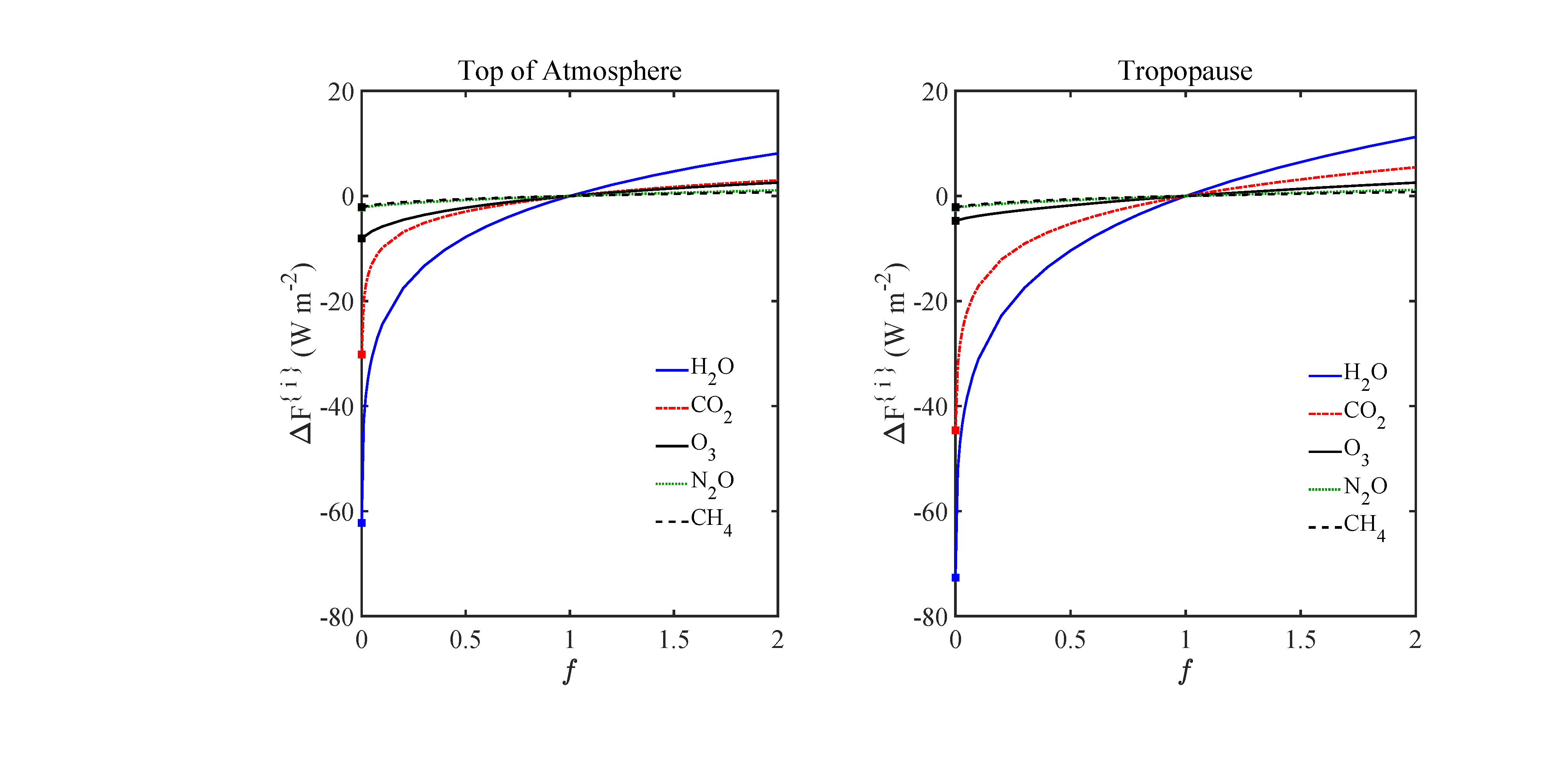}
\caption{\it The forcing increments $\Delta F^{\{i\}}$ of the five most important greenhouse gases, H$_2$O, CO$_2$, O$_3$, N$_2$O and  CH$_4$, if their current concentrations are multiplied by the factor $f$, but all other atmospheric conditions remain the same as for Fig. \ref{GGNTa}. All of the forcing increments show ``saturation,'' an initial very rapid increase for near-zero concentrations ($f\approx 0$), and an increasingly smaller rate of increase as the concentrations increase. At current atmospheric conditions with $f=1$, the saturation is most extreme for H$_2$O and CO$_2$. From reference\,\cite{WH1}.
\label{DFDC}}
\end{figure}
\begin{figure}[h]\centering
\includegraphics[height=100mm,width=1\columnwidth]{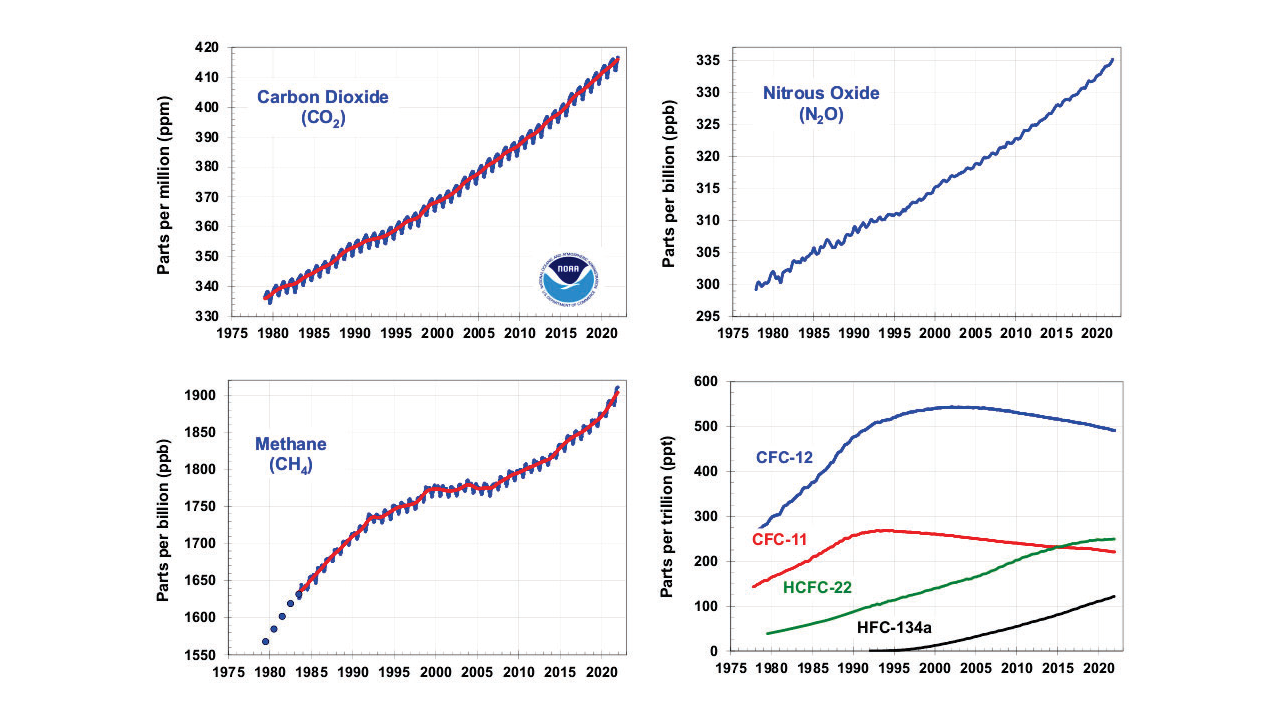}
\caption{\it Changing concentrations of Earth's main natural greenhouse gases carbon dioxide, CO$_2$, nitrous oxide, N$_2$O, methane, CH$_4$, and several halocarbon refrigerant gases. The rates of the naturally occurring greenhouse gases have suppressed zeros and are relatively slow, as shown by the century-scale doubling times $t_2$ shown in Table \ref{table4}. From reference\,\cite{dCdt}.
\label{dCdt}}
\end{figure}

\subsection{Saturation \label{sat}}
As we will discuss in more detail below, most of the increased forcing over the next half century will be from increasing concentrations $C$ of CO$_2$.  There is little debate that the forcing due to CO$_2$ is ``logarithmic.'' This means that for an increase of CO$_2$ concentrations from $C_1$ to $C_2$ the forcing will increase from $F_1$ to $F_2$, where
\begin{equation}	
F_2-F_1 =\Delta F \log_2(C_2/C_1).
\label{ff20}
\end{equation}
Here, $\log_2(x)$ denotes the base-2 logarithm of the variable $x$, for example, 
$\log_2(1)=0$, $\log_2 (2) =1$, $\log_2(4)=2$, and $\log_2(8)=3$. 
The forcing increment $\Delta F$ produced by a doubling of CO$_2$ concentrations depends on the altitude, location on Earth's surface and season of the  year. A representative number for the top of the atmosphere in a midlatitude location is $\Delta F=3.0$ W m$^{-2}$\,\cite{WH1}.  
Increasing the CO$_2$ concentrations by another 400 ppm, from $C_2 = 800$ ppm to $C_3 = 1200$ ppm would increase the forcing by
\begin{equation}	
F_3-F_2 =\Delta F \log_2(C_3/C_2)= \Delta F \log_2(3/2) = \Delta F\times 0.585. 
\label{ff24}
\end{equation}
The second 400 ppm increase of CO$_2$ concentration only increases the forcing by about 59\% of the first 400 ppm increase. This ``saturation'' of the forcing from more CO$_2$ can be seen in Fig.~\ref{DFDC}. The curves of forcing versus greenhouse gas concentration droop downward, especially for CO$_2$ and H$_2$O, and to a lesser extent, for CH$_4$ and N$_2$O.   There is a ``law of diminishing returns'' for forcing increases caused by increasing concentrations of greenhouse gases.
\begin{table}
\begin{center}
\begin{tabular}{||c c c c c c c ||}
 \hline
 Molecule &  $\hat C_1^{\{i\}}$& $d\hat C_1^{\{i\}}/dt$&$t_2^{\{i\}}$&$P^{\{i\}}$ &$dF^{\{i\}}/dt$& $\partial T^{\{i\}}/\partial t$ \\ 
 &&y$^{-1}$&y&W &W m$^{-2}$ y$^{-1}$&  K y$^{-1}$ \\ 
 \hline\hline
 CO$_2$ &$4.1\times 10^{-4}$&$2.5\times 10^{-6}$&164&$9.0 \times 10^{-26}$&$4.8\times 10^{-2}$& $8.5\times 10^{-3}$ \\  \hline
 CH$_4$ &$1.9\times 10^{-6}$&$8\times 10^{-9}$&238& $2.8\times 10^{-24}$&$4.8\times 10^{-3}$&$8.5\times 10^{-4}$  \\  \hline
N$_2$O &$3.4\times 10^{-7}$&$8\times 10^{-10}$&425&$2.1\times 10^{-23}$& $3.6\times 10^{-3}$& $6.4\times 10^{-4}$\\  \hline
\end{tabular}
\end{center}
\caption{\it Some properties of the three most important greenhouse gases (after water vapor):  carbon dioxide, methane,  and nitrous oxide. 
Altitude-averaged concentrations, $\hat C_1^{\{i\}}$, for the year 2021, and their rates of increase, $d\hat C_1^{\{i\}}/dt$ , can be found by inspection of Fig. \ref{dCdt}, and are shown in the second and third columns. Doubling times from (\ref{ff9}) are listed in the fourth column. The forcing powers per molecule from (\ref{cm14}) are displayed in the fifth column. The rates of increase of the forcing, $dF^{\{i\}}/dt$, calculated from (\ref{ff4}) for a cloud-free atmosphere, are shown in the sixth column.  The final column contains the warming rates that follow from (\ref{ff26b}), from the forcing rates of the previous column and from (\ref{ff27}).
\label{table4}}
\end{table}
\section{Future Forcings}
In Fig. \ref{DFDC} we show the forcing increments, 
\begin{equation}	
\Delta F^{\{i\}}_f=F^{\{i\}}(\hat C^{\{i\}}_f )-F^{\{i\}}(\hat C^{\{i\}}_1),
\label{cm2}
\end{equation}
caused by increasing the current concentration $C^{\{i\}}_1(z)$ of the $i$th greenhouse gas, and its altitude average 
\begin{equation}	
\hat C^{\{i\}}_1=\frac{1}{\hat N}\int_0^{\infty} dz\, C^{\{i\}}_1(z) N(z)
\label{cm6}
\end{equation}
by a factor $f$ to 

\begin{equation}	
C^{\{i\}}_f=f C^{\{i\}}_1,\quad \hbox{and}\quad \hat C^{\{i\}}_f=f \hat C^{\{i\}}_1.
\label{cm4}
\end{equation}
In (\ref{cm6}), $N(z)$ is the number density of all air molecules at the altitude $z$. The column density $\hat N$ of all atmospheric molecules, for dry air at a sea-level  pressure of 1013 mb, is
\begin{equation}	
\hat N=\int_0^{\infty}dz N(z) = 2.15 \times 10^{29} \hbox{ m}^{-2}.
\label{cm8}
\end{equation}
The rapid decrease of $N(z)$ with increasing altitude $z$ is shown in Fig. \ref{GGNTa}.

In calculating the forcing $F^{\{i\}}(\hat C^{\{i\}}_f )$ of (\ref{cm2}) we assume no change of the atmospheric temperature profile $T(z)$ or concentrations $ C^{\{j\}}_1(z)$  of other greenhouse gases ($j\ne i$). The second and third columns of Table \ref{table4}  show the  average concentrations, $\hat C^{\{i\}}_1$  in the year 2021, and the rates of growth, $d\hat C^{\{i\}}_1/dt$, both estimated by inspection of Fig.  \ref{dCdt}. 
The fourth column of Table \ref{table4}  shows the doubling times of the greenhouse gases if current rates of growth remain unchanged
\begin{equation}	
t_2^{\{i\}} =\frac{\hat C_1^{\{i\}}}{d\hat C_1^{\{i\}}/dt}.
\label{ff9}
\end{equation}
The ``partial forcings''  $F^{\{i\}}_f$  of (\ref{cm2}) increase with increasing average concentrations $\hat C^{\{i\}}_f$ at the rate
\begin{equation}	
\frac{\partial F^{\{i\}}_f}{\partial \hat C^{\{i\}}_f}=\frac{1}{\hat C^{\{i\}}_1}\frac{d}{df}\Delta F^{\{i\}}_1,
\label{cm10}
\end{equation}
where the incremental forcings $\Delta F^{\{i\}}_f=\Delta F^{\{i\}}$ were shown in Fig. \ref{DFDC}.
Partial derivative symbols, $\partial$, are used on the left of (\ref{cm10}) as a reminder that only the concentration of the $i$th greenhouse gas is varied from the conditions of Fig. \ref{GGNTa}.  The column density $\hat N^{\{i\}}$ of the greenhouse gas  $i$ is related to its altitude-averaged concentration by
\begin{equation}	
\hat N^{\{i\}}=\hat N \hat C^{\{i\}}.
\label{cm12}
\end{equation}
Multiplying both sides of (\ref{cm10}) by $1/\hat N$ we see that the rate of increase of forcing with increasing column density $\hat N^{\{i\}}$ of the greenhouse gas $i$ is
\begin{equation}	
P^{\{i\}}=\frac{\partial F^{\{i\}}}{\partial \hat N^{\{i\}}}=\frac{1}{\hat N \hat C^{\{i\}}_1}\frac{d}{df}\Delta F^{\{i\}}.
\label{cm14}
\end{equation}
The unit of (\ref{cm14}) is watt so we can  think of the quantity $P^{\{i\}}$ as the additional forcing produced by 1 molecule per square meter of the greenhouse gas $i$ above the Earth's surface.  Values of $P^{\{i\}}$, that follow from data like that of Fig. \ref{DFDC} and (\ref{cm14}), are entered in the fifth column of Table \ref{table4}.
The values of $P^{\{i\}}$ depend  on altitude, as one can see from Fig. \ref{DFDC}. Those listed in Table \ref{table4} are for the tropopause altitude (11 km) of a representative temperate latitude.

For current concentrations the per-molecule forcing power of methane $P^{\{{\rm CH}_4\}}$ is 31 times larger than that of carbon dioxide, $P^{\{{\rm CO_2}\}}$, and the per-molecule forcing power of nitrous oxide, $P^{\{{\rm N_2 O}\}}$, is 233  times larger. This is mostly due to the ``saturation" of the absorption bands of CO$_2$ that was discussed in Section \ref{sat}.
The current density of CO$_2$ molecules is about 220 times greater than that of CH$_4$ molecules and about 1300 times greater than that of N$_2$O. So, the absorption bands of CO$_2$ are much more saturated than those of CH$_4$ or N$_2$O.  Reference \,\cite{WH1} shows that in the dilute limit, where molecules do not absorb each other's radiation before it can escape to space, CO$_2$ is the most potent greenhouse gas molecule. Each additional CO$_2$ molecule in the dilute limit causes about 5.3, 1.6 or 1.8  times more tropospheric forcing increase than an additional molecule of CH$_4$, N$_2$O, or H$_2$O.

Since the column density of the greenhouse gas $i$ increases at the rate $d\hat N^{\{i\}}/dt = \hat N d\hat C^{\{i\}}/dt$, the rate of increase of the forcing is
\begin{equation}	
\frac{d F^{\{i\}}}{dt} =\hat N P^{\{i\}} \frac{d\hat C^{\{i\}}}{dt}.
\label{ff4}
\end{equation}
The values of (\ref{ff4}) are listed in the sixth column of Table \ref{table4}.
The total rate of increase in forcing is
\begin{equation}	
\frac{dF}{dt}=\sum_{i}\frac{d F^{\{i\}}}{dt}.
\label{ff6}
\end{equation}
Using data from Table \ref{table4}, we see that the total rate of increase of forcing, due to increasing concentrations of CO$_2$, CH$_4$ and N$_2$O, is
\begin{equation}	
\frac{dF}{dt}=0.056 \hbox{ W m$^{-2}$ y$^{-1}$}.
\label{ff6a}
\end{equation}
This is a representative number for a midlatitude location. Analogous numbers would have to be averaged over  Earth's surface and over a year to be useful for ``global'' estimates.

\begin{figure}[h]\centering
\includegraphics[height=100mm,width=1\columnwidth]{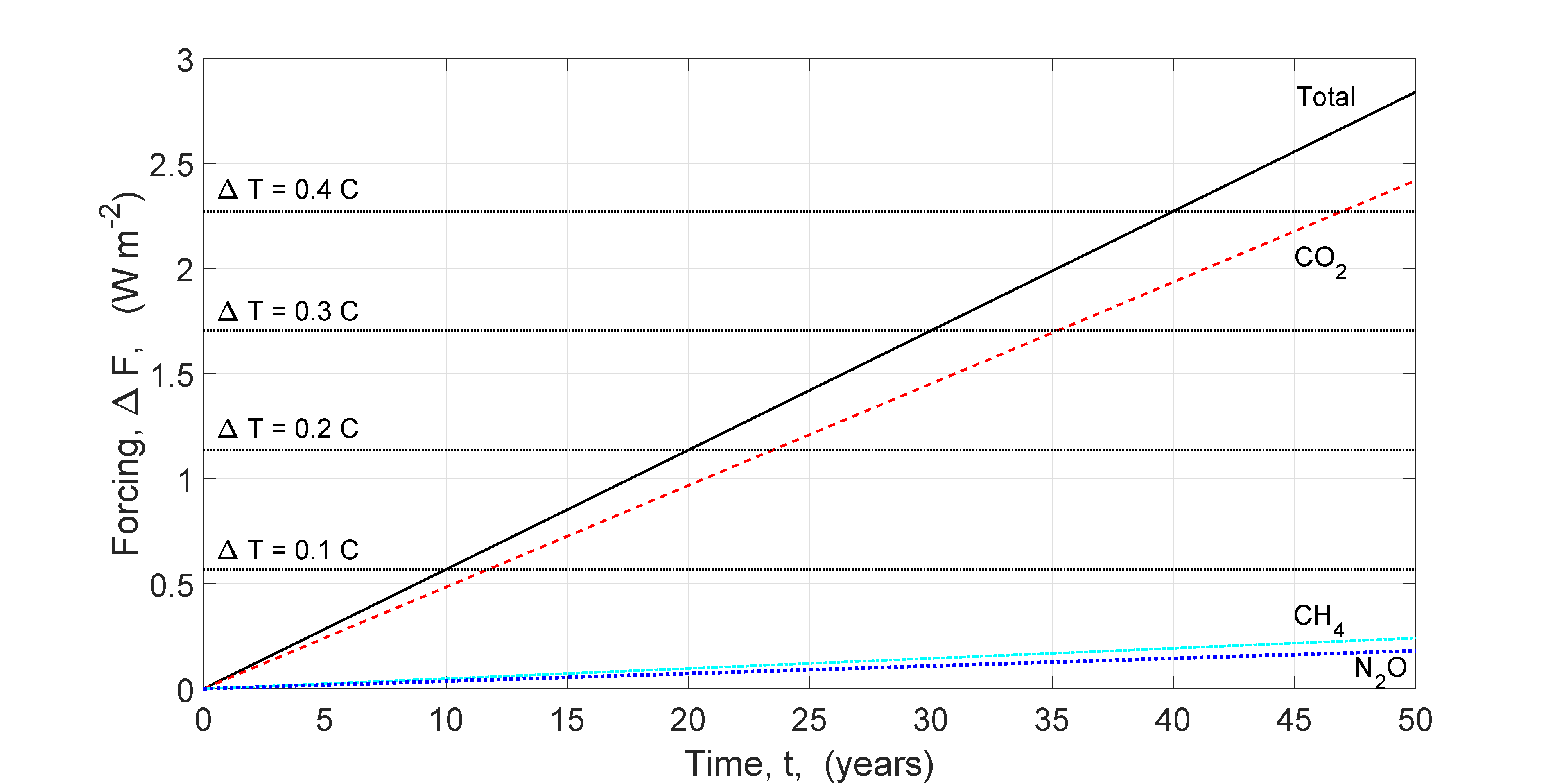}
\caption{\it Projected midlatitude forcing increments at the tropopause from continued increases of CO$_2$, CH$_4$ and N$_2$O at the rates of
Fig. \ref{dCdt}  for the next 50 years. The projected forcings are very small compared to the current tropospheric forcing of 137 W m$^{-2}$. The temperature-increase contours correspond to the average warming of the past few decades, $0.01$  C y$^{-1}$. The total forcing rate (warming rate) is approximately 85\% from CO$_2$, 9\% from CH$_4$ and 6\% from N$_2$O. 
\label{dFdt3}}
\end{figure}

\subsection{Temperature changes due to forcing changes}
Instantaneous forcing changes due to instantaneous changes in the concentrations of greenhouse gases, but with no other changes to the atmosphere, can be calculated accurately, as we have outlined above. 
The next step, using instantaneous forcing changes to predict changes in Earth's temperatures,  is much  harder.  The very concept of ``Earth's temperature'' is not well defined. Earth's temperatures are highly variable in time, in latitude, in altitude, or depth in the ocean. The Earth does not have a single temperature. Even defining a sensible average temperature or temperature anomaly is problematic.  Lindzen and Christy\,\cite{LC} have given a good discussion of how elusive the concept of Earth's ``temperature'' is.

Suppose that the average solar heating of the Earth were to remain the same after an instantaneous doubling in the concentration of  CO$_2$. Since the additional greenhouse gas has slightly decreased the  thermal radiation to space (or radiative cooling), the Earth would begin to accumulate heat at the rate of a few Watts per square meter. The unbalanced addition of solar heat would continue until  the conditions of the atmosphere, the surface and the oceans change enough to equalize the average values of solar heating and radiant cooling.   Emission of thermal radiation increases rapidly with increasing temperature so a small amount of warming of the atmosphere or the surfaces of the land and oceans can increase radiative cooling.  An increase in  cloud cover can reflect more solar energy back into space before it is converted to heat.  The details of how balance is restored  are extremely complicated because of the complexity of Earth's climate physics. The atmosphere and oceans convect large amounts of heat from the tropics to the poles, so more radiation is emitted from polar regions than the solar heating there. Cloud cover and ice extent will probably change as well. But because the relative changes in radiative forcing are so small, between one and two percent,  as one can see from Fig. \ref{ZzCMa}, minor changes of the present climate will be needed to reestablish balance.   

For the sake of further discussion,
we  assume that some average temperature or temperature anomaly $T$ of the Earth can be reliably defined and measured. Then we can write the dependence of this temperature $T$ on time $t$ as
\begin{equation}	
\frac{dT}{dt} =\frac{\partial T^{\{\rm nat\}}}{\partial t}+\frac{\partial T^{\{\rm gg\}}}{\partial t}.
\label{ff22}
\end{equation}
The first term on the right of (\ref{ff22}),  $\partial T^{\{\rm nat\}}/\partial t$, is the temperature change due to natural causes. The geological history of Earth clearly shows that the natural variations of temperature can be large on all time scales, from hours to millennia and longer. The second term, $\partial T^{\{\rm gg\}}/\partial t$, is the rate of change due to increasing concentrations of greenhouse gases.

Temperature records have been available at many sites since the late 1800s. So, the value of $dT/dt$ on the left side of the equation (\ref{ff22}) is known, at least approximately.  But early temperature readings were not taken for most parts of the Earth's surface, including the oceans, the polar regions,  the Sahara Desert, the Amazon Basin, etc. So, the observed $dT/dt$ has especially large uncertainties before the year 1900.  According to the IPCC, surface temperatures have increased by about 1.1 K\,\cite{Forster}  since the year 1850.   The observed warming exhibits significant decadal fluctuations.  The average surface temperature increased by about 0.5 K during 1900 - 1940, remained relatively constant from 1940 to 1980, increased from 1980 to 2000 and, surprisingly, leveled off for over a decade after 2000. This is known as the global warming hiatus.  Since then, an overall warming trend may have resumed, although satellite measurements indicate that another pause may be underway\,\cite{Spencer}.

Since the year 1979, satellite measurements of  thermal millimeter (mm) waves emitted by O$_2$ molecules in the atmosphere have been used to infer temperatures in Earth's atmosphere\,\cite{MSU,Spencer}. For mm-waves near 60 GHz in frequency, there is a very strong absorption band due to spin-flip transitions in the O$_2$ molecule.  O$_2$ is nearly transparent to visible sunlight and to most of the thermal radiation from the Earth, but it does absorb and thermally emit mm waves.  By tuning the mm-wave radiometers on satellites to appropriate frequencies near 60 GHz, one can measure radiation intensity (and the corresponding temperature)  from relatively narrow ranges of atmospheric altitude, from the lower troposphere to the stratosphere. Since satellite measurements began in the year 1979, the warming rate of ``the global lower atmosphere'' has averaged about 0.015  K y$^{-1}$\,\cite{Spencer}. 

It is hard to construct a climate model that does not predict faster warming of the troposphere than the surface\,\cite{WH1}. The moist adiabats (or pseudoadiabats) that give a reasonably accurate fit to the average temperature profiles measured by radiosondes on weather balloons, do not have simple linear lapse rates of 6.5 K km$^{-1}$. Instead, they ``bend'' in such a way that higher altitudes warm more than lower altitudes\,\cite{WH1}.  The different warming rates are due to differences in how moisture condenses to liquid water or ice crystals and releases latent heat at various altitudes. The mm-wave emissivity of the surface is so variable and uncertain that measurements of surface temperatures from satellites are substantially less accurate than measurements of atmospheric temperatures at higher altitudes. That satellites observe faster atmospheric warming rates than surface warming rates measured by ground stations may be due to the effects of moist adiabats. 
  
No one knows how much of the observed surface or atmospheric warming is due to natural causes, the first term on the right side of (\ref{ff22}), and how much is due to increasing concentrations of greenhouse gases, the second term. Some temperature increase over the past two centuries must have been a natural recovery from the Little Ice Age. For the sake of argument, we will assume that in recent decades the amount of  temperature increase that can be solely ascribed to greenhouse gases amounts to
\begin{equation}	
\frac{d T^{\{\rm gg\}}}{d t}= 0.01 \hbox{  K y}^{-1},
\label{ff24}
\end{equation}
or 1 C per century.   The relative contributions of CO$_2$, N$_2$O and CH$_4$ to the warming are determined by the relative forcings, that can be accurately calculated. The relative contributions are independent of the value of (\ref{ff24}).

The thermal flux changes  (the negatives of the forcing changes)  at the top of the atmosphere are about 1\% of the flux to space for doubling CO$_2$, as shown in Fig. \ref{ZzCMa}. And these forcing changes occur very slowly.  From Table \ref{table4} we see that doubling CO$_2$ at current rates would require about 160 years. It is therefore a good assumption that the part of the rate of temperature change in (\ref{ff22}) that is due to greenhouse gases is proportional to the rate of forcing change, and we can write 
the contribution of the $i$th greenhouse gas to the warming rate as
\begin{equation}	
\frac{\partial T^{\{i\}}}{\partial t}=\left(\frac{\partial T}{\partial Z}\right)\frac{dF^{\{i\}}}{dt}.
\label{ff26b}
\end{equation}
The total warming rate due to increases of all greenhouse gases is then
\begin{eqnarray}	
\frac{dT^{\{\rm gg\}}}{dt}&=&\sum_i \frac{\partial T^{\{i\}}}{\partial t}\nonumber\\
&=&\left(\frac{\partial T}{\partial Z}\right)\frac{dF}{dt}.
\label{ff26a}
\end{eqnarray}
Substituting (\ref{ff6a})  and (\ref{ff24}) into (\ref{ff26a}), we find
\begin{equation}	
\frac{\partial Z}{\partial T}=\left(\frac{\partial T}{\partial Z}\right)^{-1}
= 5.6 \hbox{ W m$^{-2}$ K$^{-1}$}.
\label{ff27}
\end{equation}
Here, we have assumed that  the  value $dF/dt$ of (\ref{ff6a}), computed for the midlatitude tropopause of clear skies, is an adequate metric for the complicated range
of values of  $dF/dt$ that characterize other latitudes, altitudes and cloud conditions.
Using (\ref{ff27}) and values of $dF^{\{i\}}/dt$ from Table \ref{table4} in (\ref{ff26b}) we find the warming rates $\partial T^{\{i\}}/\partial t$ due to the $i$th species of greenhouse gases. The values are listed in the last column of Table \ref{table4}.

Equation (\ref{ff27}) is very important.  It says that the average global temperature, $T$,  for all its logical problems\,\cite{LC}, is relatively insensitive to changes in the concentration of greenhouse gases. A forcing of 5.6 W m$^{-2}$ is needed to increase $T$ by 1 K or 1 C.  Instantaneously doubling the concentration of CO$_2$, which will take more than a century according to Table \ref{table4}, only decreases the flux by  5.5  W m$^{-2}$ at the tropopause (and only by 3.0 W m$^{-2}$ at the top of the atmosphere)\,\cite{WH1}.  So, for the estimated value (\ref{ff27}) of $\partial Z/\partial T$, a warming of 0.98 K would increase the tropopause flux back to the value it had before doubling CO$_2$ concentrations.

To get some feeling for how physically reasonable (\ref{ff27}) is, we note that
if the Earth radiated as a blackbody, in accordance with the Stefan-Boltzmann law (\ref{rff3}), at a temperature $T = 288.7 \hbox{ K}$, the rate of change of flux with temperature would be
\begin{equation}	
\frac{\partial Z}{\partial T} =
4\sigma T^3 =5.5 \hbox{ W m$^{-2}$ K$^{-1}$}. 
\label{ff28}
\end{equation}
The estimates (\ref{ff27}) and (\ref{ff28}) are within a few percent of each other.
In reference \,\cite{WH1}  a representative rate of decrease of flux at the top of a cloud-free atmosphere due to a uniform increase  of temperatures at all altitudes was calculated to be
\begin{equation}	
\frac{\partial Z}{\partial T} =3.9 \hbox{ W m$^{-2}$ K$^{-1}$}.
\label{ff32}
\end{equation}
This is about 30\% smaller than (\ref{ff27}).

The elaborate computer models of the IPCC predict that much larger changes of temperature are needed to increase the cooling flux and compensate for the forcing of greenhouse gases. In the year 2021, the IPCC\,\cite{Forster} estimated that doubling CO$_2$ concentrations would cause a  ``most likely'' warming  of 
\begin{equation}	
\partial T= 3.0 \hbox{ K}.
\label{ff34}
\end{equation}
Taking the flux increase  to be
\begin{equation}	
\partial Z = 5.5 \hbox{ W m$^{-2}$ K$^{-1}$},
\label{ff36}
\end{equation}
as calculated in reference\,\cite{WH1}  for the tropopause,
we find
\begin{equation}	
\frac{\partial Z}{\partial T} =1.8 \hbox{ W m$^{-2}$ K$^{-1}$}. 
\label{ff38}
\end{equation}
If  $\partial Z/\partial T$ were really as small as (\ref{ff38}), the warming rate  (\ref{ff26a}) of the past few  decades would have been
\begin{equation}	
\frac{\partial T^{\{\rm gg\}}}{\partial t} = 0.031 \hbox{ K y}^{-1}
\label{ff40}
\end{equation}
three times larger than the observed rate, $0.010 \hbox{ K y}^{-1}$, of   (\ref{ff24}). 
\begin{figure}[h]\centering
\includegraphics[height=100mm,width=1\columnwidth]{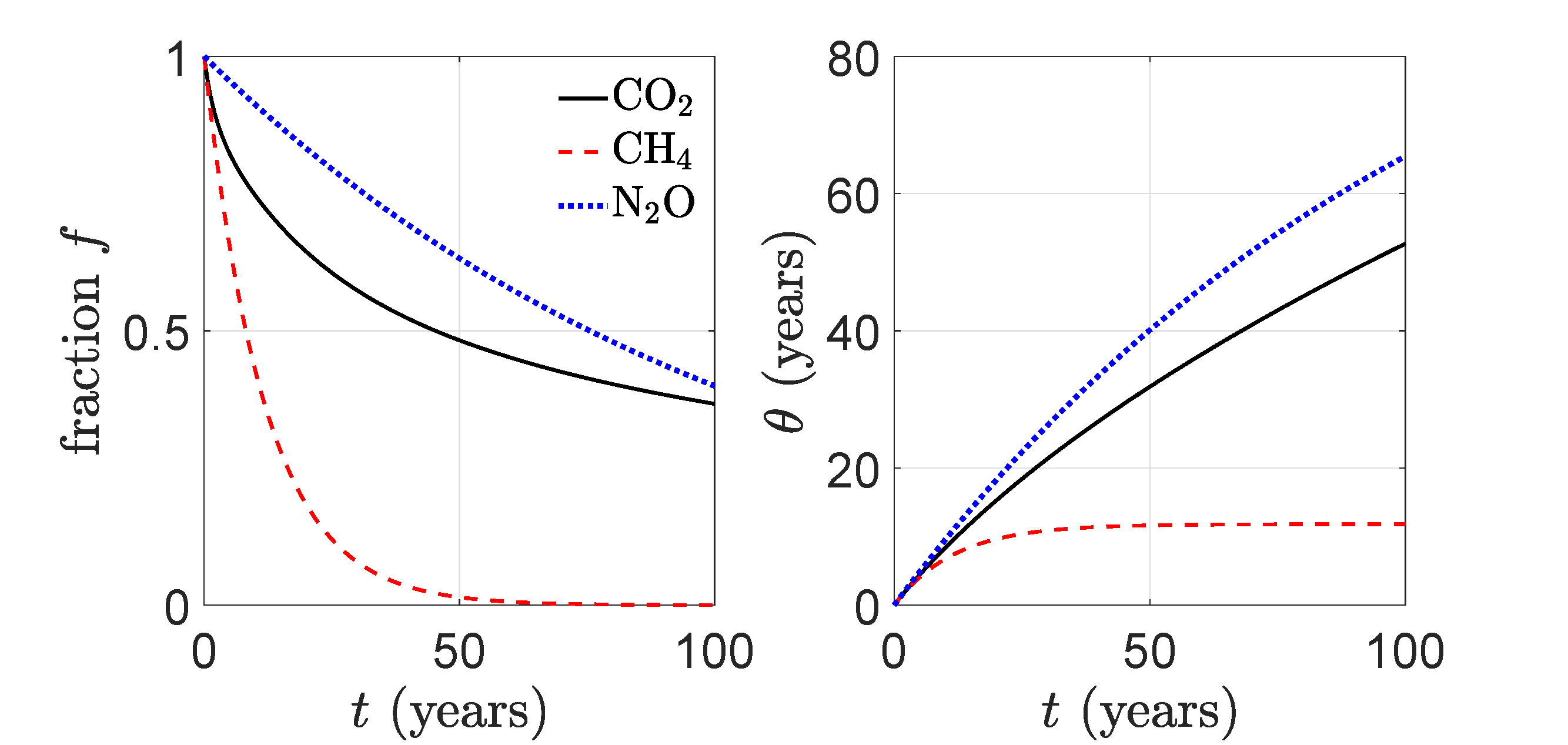}
\caption{\it {\bf Left:} Fractions $f = f(t)$ of greenhouse gases remaining in the atmosphere at a time $t$ after one unit mass of gas was injected at time $t=0$.  The fractions are described by the decaying exponentials of (\ref{gwp8}).  {\bf Right:} Forcing times $\theta=\theta(t)$ for observation-time intervals $t$. These are described by (\ref{gwp10}).
\label{gwp}}
\end{figure}

\subsection{Global warming potential}
The powers per molecule $P^{\{i\}}$ of (\ref{cm14}) measure the relative forcings of different greenhouse molecules. They play a similar role to the {global warming potential} (GWP), that is defined in  Section 8.7.1.2 of the 2018 IPCC document \,\cite{IPCC}, in terms of radiative forcings (RF), by:
\begin{quote}
The Global Warming Potential (GWP) is defined as the time-integrated RF due to a pulse emission of a given component, relative to a pulse emission of an equal mass of CO$_2$ (Figure 8.28a and formula). The GWP was presented in the First IPCC Assessment (Houghton et al., 1990), stating ``It must be stressed that there is no universally accepted methodology for combining all the relevant factors into a single global warming potential for greenhouse gas emissions. A simple approach has been adopted here to illustrate the difficulties inherent in the concept, ...''. Further, the First IPCC Assessment gave no clear physical interpretation of the GWP.
\end{quote}
To translate these words into equations that allow calculations of global warming potentials,   
we denote the molecular mass of a greenhouse molecule of type $i$ by $m^{\{i\}}$ and the {\it forcing time} by $\theta^{\{i\}}$. Then we can quantify the IPCC definition above by writing the global warming potential of the $i$th type of greenhouse gas as
\begin{equation}	
\hbox{GWP}^{\{i\}} = \frac{\langle \hbox{RF}^{\{i\}}\rangle}{\langle \hbox{RF}^{\{{\rm CO}_2\}}\rangle}.
\label{gwp2}
\end{equation}
The  ``time-integrated RF''  (time-integrated radiative forcing) per unit mass of the $i$th greenhouse gas is
\begin{equation}	
\langle \hbox{RF}^{\{i\}}\rangle=\frac{P^{\{i\}}\theta^{\{i\}}}{m^{\{i\}}}.
\label{gwp4}
\end{equation}
The units of $\langle \hbox{RF}^{\{i\}}\rangle$ are joules per square meter per unit mass. The  $\langle \hbox{RF}^{\{i\}}\rangle$ can be thought of as the amount of ``forcing heat" acquired by a square-meter column of the atmosphere during the observation time $t$ after the pulse emission of one unit mass of greenhouse gas of type $i$.
The forcing  time $\theta^{\{i\}}$ for an observation  time interval $t$ is
\begin{equation}	
\theta^{\{i\}}=\int_0^t dt' f^{\{i\}}(t').
\label{gwp6}
\end{equation}
Here, $f^{\{i\}}(t')$ is the fraction of the excess greenhouse gas molecules remaining in the atmosphere at a time $t'$ after the ``pulse emission.''  Excess greenhouse molecules can remain in the atmosphere long after the greenhouse molecules of the original pulse have exchanged with the land and oceans because they are replaced with equivalent molecules that continue to provide radiative forcing. An example is the relatively short time needed for  the exchange of $^{14}$CO$_2$ molecules from atmospheric  nuclear weapon tests with CO$_2$ molecules of the ocean. It takes a much longer time for excess CO$_2$ concentrations to decay away. The excess atmospheric concentration of molecules,  caused by the emitted pulse of the species $i$, can be removed from the atmosphere by various mechanisms. For example, CO$_2$ is absorbed by the biosphere and oceans.  CH$_4$ is oxidized by OH radicals and other atmospheric gases. N$_2$O is photodissociated by solar ultraviolet radiation in the stratosphere, etc.  

For a  global warming potential to make physical sense, there must be some equilibrium concentration of the greenhouse gas of interest for which natural sources and sinks are in balance, and for which the concentration would  not change with time if there were no human influences.
Though plausible, there is no compelling observational evidence for the existence of equilibrium concentrations of any greenhouse gas. Indeed, ice core records show large variations of CO$_2$ concentrations over  periods of hundreds of thousands of years, especially between glacial maxima and interglacials of the current ice age.  The variations of CO$_2$ appear to have been driven by variations of Earth's average temperature, as inferred from variations of the stable isotope $^{18}$O  in the ice surrounding the trapped air bubbles.   Changes in  ice-core concentrations of CO$_2$  follow changes in temperature by several centuries\,\cite{Petit}.

\begin{table}
\begin{center}
\begin{tabular}{||c | c c c | c c c  ||}
 \hline
 Molecule & $P^{\{i\}}$ &$P^{\{i\}}/P^{\{{\rm CO}_2\}}$&$m^{\{i\}}$&\multicolumn{3}{c||}{This Work} 
 \\ $i$&W &&amu &GWP$^{\{i\}}$(0) &GWP$^{\{i\}}$(20) &GWP$^{\{i\}}$(100)\\ 
 \hline\hline
 CO$_2$ &$9.0 \times 10^{-26}$&1&44&1& 1&1\\ \hline
 CH$_4$ &$2.8\times 10^{-24}$&31&16& 85.5&53.9&19.2 \\ \hline
N$_2$O &$2.1\times 10^{-23}$&233&44& 233&279 &290\\ \hline
\end{tabular}
\end{center}
\caption{\it Forcing powers per molecule, $P^{\{i\}}$ from (\ref{cm14}), global warming potentials GWP$^{\{i\}}(t)$ from (\ref{gwp12}) for observation times  $t\to 0$ years,  $t=20$ years and $t= 100$ years.  
\label{table3}}
\end{table}
\begin{table}
\begin{center}
\begin{tabular}{||c | c c  ||}
 \hline
 Molecule &\multicolumn{2}{c||}{IPCC 2022} \\
 $i$&GWP$^{\{i\}}$(20) &GWP$^{\{i\}}$(100)\\ 
 \hline\hline
 CO$_2$ &1&1\\ \hline
 CH$_4$ &$82.5\pm 25.8$&$29.8\pm 11$ \\ \hline
N$_2$O &$273\pm 118$&$273\pm 130$\\
\hline
\end{tabular}
\end{center}
\caption{\it Global warming potentials  for fossil-fuel methane and nitrous oxide from the 2021 IPCC report\,\cite{IPCC}. The observation times are $t=20$ years and $t= 100$ years.
\label{table5}}
\end{table}

To facilitate further discussion, we will assume that equilibrium concentrations exist. In that  case,
it is customary and convenient to describe the excess fraction $f^{\{i\}}(t)=f(t)$  with a sum of $n+1$ exponentially decaying components, of amplitudes $a_j$ and time constants $\tau_j$\,\cite{Harvey, Joos}
\begin{equation}	
f(t) =\sum_{j=0}^n a_j e^{-t/\tau_j},\quad\hbox{where}\quad \sum_{j=0}^n a_j = 1.
\label{gwp8}
\end{equation}
Substituting (\ref{gwp8}) into (\ref{gwp6}) we find that the forcing time, $\theta^{\{i\}}=\theta$, is
\begin{equation}	
\theta=\sum_{j=0}^n a_j\tau_j\left(1-e^{-t/\tau_j}\right).
\label{gwp10}
\end{equation}
For the limiting case of infinitely long time constants, $\tau_j\to\infty$, one should make the replacement $\tau_j\left(1-e^{-t/\tau_j}\right)\to t$ in (\ref{gwp10}). From inspection of (\ref{gwp10}) one can conclude that
the forcing time $\theta^{\{i\}}$ is always less than or equal to the observation time, $\theta^{\{i\}}(t)\le t$.
Using (\ref{gwp10}) in (\ref{gwp2}) we find that the global warming potentials are
\begin{equation}	
\hbox{GWP}^{\{i\}}= \left(\frac{P^{\{i\}}}{P^{\{{\rm CO}_2\}}}\right)\left(\frac{m^{\{{\rm CO}_2\}}}
{m^{\{i\}}}\right)\left(\frac{\theta^{\{i\}}}{\theta^{\{{\rm CO}_2\}}}\right).
\label{gwp12}
\end{equation}

Unlike the forcing powers $P^{\{i\}}$ per greenhouse molecule of type $i$, which can be accurately calculated, the time dependence of the fractions $f^{\{i\}}(t)$ is not well known. So, the global warming potentials GWP$^{\{i\}}$ are of limited quantitative value. To add further uncertainty, ``indirect effects" are sometimes included in the GWP$^{\{i\}}$, for example, the effects of the CO$_2$ and H$_2$O molecules which result from the oxidation of CH$_4$.    
For this paper we will consider only the direct forcing of the greenhouse gases. 

For CH$_4$ and N$_2$O,
Table 7.15 of Chapter 7 of the IPCC report for the year 2021\,\cite{Forster} gives lifetimes for
single-exponential versions of (\ref{gwp8}) 
\begin{equation}	
\tau^{\rm CH_4}=11.8\pm 1.8\hbox{ y}\quad\hbox{and}\quad \tau^{ \rm N_2O}=109\pm 10\hbox{ y}.
\label{gwp14}
\end{equation}
For CO$_2$, Harvey\,\cite{Harvey} uses a five-exponential form of (\ref{gwp8}) with the parameters
\begin{equation}
\left[\begin{array}{c}a_0\\ a_1\\
a_2\\ a_3 \\ a_4\end{array}\right]
=\left[\begin{array}{c}.131\\ .201\\ .321\\ .249 \\.098\end{array}\right]\quad \hbox{and}\quad
\left[\begin{array}{c}\tau_0\\ \tau_1\\
\tau_2\\ \tau_3 \\ \tau_4\end{array}\right]
=\left[\begin{array}{c}\infty\\ 362.9\\ 73.6\\ 17.3 \\ 1.9\end{array}\right]\hbox{ y}.
\label{gwp16}
\end{equation}
Alternate parameterizations of $f^{\{i\}}$ for CO$_2$, for example, those of Joos {\it et al.}\,\cite{Joos}, give GWP's that do not differ more than the uncertainties of the estimates. 

The left panel of 
Fig. \ref{gwp} shows excess fractions $f^{\{i\}}$ of CO{$_2$, CH$_4$ and N$_2$O, calculated with (\ref{gwp8}) and with the parameters of (\ref{gwp14}) and (\ref{gwp16}). The right panel of Fig. \ref{gwp} shows the forcing times $\theta^{\{i\}}$ of (\ref{gwp10}). Table \ref{table3} shows representative global warming potentials calculated with (\ref{gwp12}) and the parameters of (\ref{gwp14}) and (\ref{gwp16}). These are consistent with those of the 2021 IPCC report\,\cite{Forster} shown in Table \ref{table5}.  The somewhat larger values of GWP$^{\{\rm CH_4\}}$ may  be due to indirect effects included in the IPCC calculations.
\section{Nitrogen in the Biosphere  \label{bio}}
Nitrogen, the main focus of this paper, is the third most abundant material, after water and carbon dioxide, needed for plant growth.   The ultimate source of nitrogen in the biosphere is elemental diatomic nitrogen molecules, N$_2$, from the atmosphere. The weathered rocks from which soils are formed contain negligible amounts of nitrogen. How atmospheric nitrogen is converted into forms that can be used by living organisms and ultimately cycled back into the atmosphere is called the {\it nitrogen cycle}\,\cite{Bernhard}.  Fig. \ref{Ncycle} gives a simplified sketch of the nitrogen cycle. Most soil nitrogen is eventually returned to the atmosphere as diatomic nitrogen molecules, N$_2$, in the process of {\it denitrification}.  But a few percent is returned as the greenhouse gas molecules nitrous oxide or N$_2$O.  The biosphere is the main source of N$_2$O in the atmosphere.
\begin{figure}[t]\centering
\includegraphics[height=100mm,width=1\columnwidth]{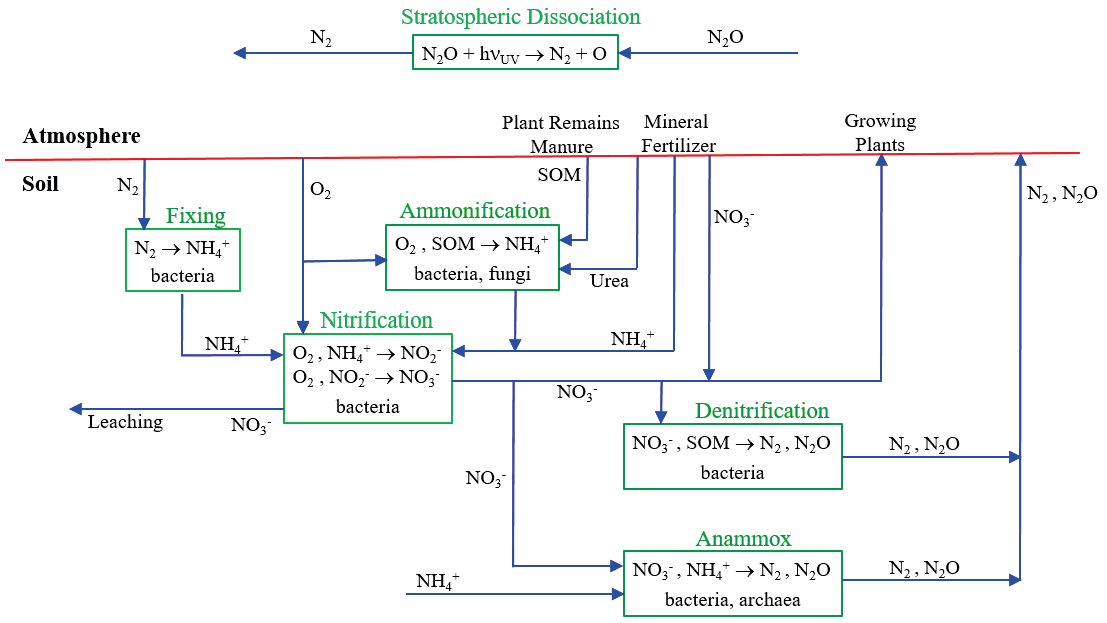}
\caption{\it Simplified nitrogen cycle for the soil. SOM is soil organic matter. Most of the fixed nitrogen of the soil is returned to the atmosphere as N$_2$ molecules, with a small fraction as N$_2$O molecules.  See the text for more details.
\label{Ncycle}}
\end{figure}
For land plants, water is the most important requirement. Without adequate water, soil fertility is irrelevant, plants would die, and there would only be barren desert. Of course, water is never in short supply for aquatic plants. 

After water, carbon dioxide is the next most essential requirement. Life would be impossible without the carbon atoms that are the main constituents of carbohydrates, hydrocarbons, proteins, nucleic acids, and the many other organic compounds of which living things are built.  The carbon in living organisms comes from carbon dioxide molecules of the atmosphere and oceans. When living organisms die, much of  the organic carbon  of their remains is eventually oxidized back to CO$_2$ by microorganisms and returned to the atmosphere and ocean.  But under special anoxic conditions, some of the organic-rich sediments can be preserved, then buried deeply and eventually converted to coal, oil and natural gas.

Current atmospheric concentrations of carbon dioxide are low by the standards of geological history\,\cite{Berner}. Insufficient atmospheric carbon dioxide is retarding plant growth\,\cite{Taylor}. The modest growth of CO$_2$ concentrations over the past century has contributed to the greening of the Earth, which can be observed from satellites\,\cite{greening} and to significant increases of crop yields\,\cite{Taylor}.

\begin{figure}[h]\centering
\includegraphics[height=100mm,width=1\columnwidth]{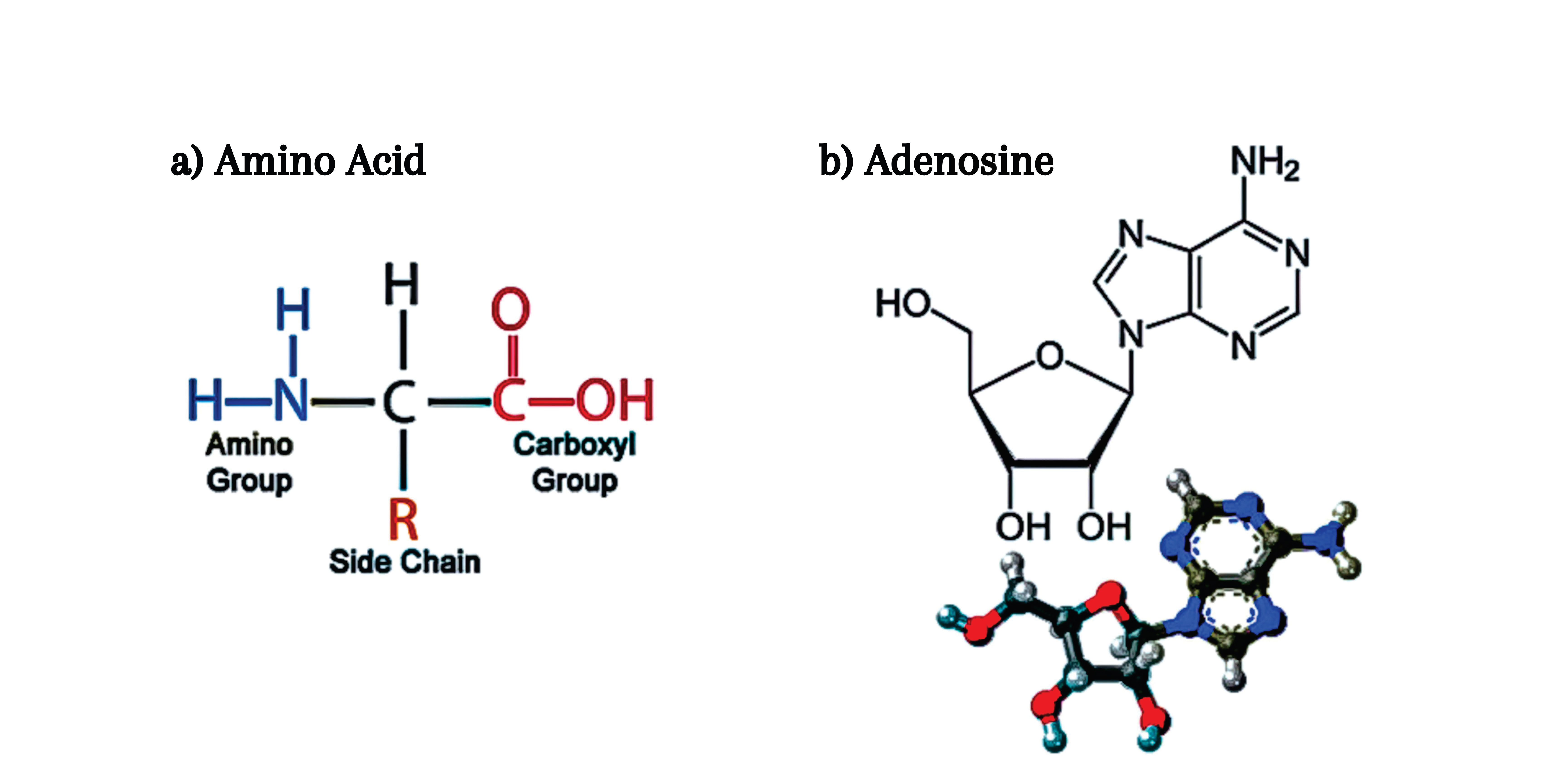}
\caption{\it {\bf Left.} An amino acid molecule, the building block of the proteins of living organisms. From reference \,\cite{amino}.  {\bf Right.} The nucleoside adenosine.  Attaching one phosphate, PO$_4^{3-}$ to the OH on the left makes a nucleotide, adenosine monophosphate, one of the four building blocks of ribonucleic acid (RNA).  Attaching a chain of three phosphates makes adenosine triphosphate (ATP), the main energy-transport molecule of life. From reference \,\cite{adenosine}.
\label{N-life}}
\end{figure}
Nitrogen, the third most essential requirement for plant growth, is a key part of proteins, the structural material of all life.
Proteins are long-chain polymers (peptide chains) of amino acids like the generic one sketched in the left side of Fig. \ref{N-life}.   Amino acids have at least one nitrogen atom in the amino group, -NH$_2$.  There are 20 different amino acids commonly found in plants and animals, each with its own unique side chain R.  Six side chains R contain additional nitrogen atoms \,\cite{aminoacid}. Though far less abundant than amino acids, there are many other nitrogen-containing molecules that are essential for life. One example is  the adenosine molecule, shown on the right side of Fig. \ref{N-life}. Other examples include the porphyrin rings that hold the iron ions of hemoglobin and cytochromes, or the magnesium ions of chlorophyll, or the B vitamins, or the genetic material of cells, deoxyribonucleic acid (DNA).

The biosynthetic machinery of plants uses ammonium ions, NH$_4^+$ as a source of nitrogen atoms in organic molecules.  When living organisms die,  microorganisms convert most of the nitrogen from their remains back into ammonium ions and nitrate ions, NO$_3^-$. This process of {\it mineralization} allows nitrogen to be used by new generations of living organisms and recycled many times through the soil. But the fixed nitrogen of the soil is eventually lost by leaching or into the atmosphere as N$_2$ or N$_2$O molecules in the process of  {\it denitrification}, which is carried out by other microorganisms.

In addition to using recycled, mineralized nitrogen, some
microorganisms of the soil and  oceans  can convert diatomic nitrogen, N$_2$, from the atmosphere into ammonium ions NH$_4^+$. {\it Fixing} nitrogen in this way is a highly energy-intensive step, since the triple bond energy of N$_2$,  about 9.8 electron volts (eV)\,\cite{bonds}, is one of the largest in nature. Within nitrogen - fixing microorganisms, nitrogenase, itself a protein containing many nitrogen atoms, catalyzes the overall reaction 
\begin{equation}
{\rm N}_2+3{\rm H}_2 +2\hbox{ H$_2$O}+ 16\, \hbox{ATP}\to 2 {\rm NH}_4^++2{\rm OH}^-+ 16\, \hbox{ADP}+16 \hbox{P}_i.
\label{na2}
\end{equation}
The hydrogen atoms  and the adenosine triphosphate (ATP) energy-carrier molecules on the left side of the equation are usually derived from the metabolism of energy-storage  carbohydrates, such as glucose, C$_6$H$_{12}$O$_6$, which is a major part of plant tissue, often in polymeric forms like starch or cellulose. Starch, which consists of single chains of glucose molecules, is easily decomposed to glucose.  Cellulose fibers, the main building material of plants, are hydrogen-bonded polymers of glucose with slightly different bond geometries between the monomers than for starch. Cellulose is much more resistant to decomposition than starch. Fungi and soil microorganisms can disassemble cellulose and release glucose.

The detailed biochemistry summarized by (\ref{na2}) has very complicated pathways, and it is a {\it tour de force} for organisms that can implement it.   Breaking  the N$_2$ bond requires the chemical energy of at least 16 ATP molecules.
The chemical energy, or Gibbs free energy,
\begin{equation}
\Delta G = \Delta H - T\Delta S,
\label{na3}
\end{equation}
available from each conversion of ATP to adenosine diphosphate (ADP) and an inorganic phosphate ion (P$_i$), depends on the absolute temperature $T$, the ionic strength, the pH, and other factors. It involves not only the enthalpy increment $\Delta H$  from making or breaking chemical bonds but also large entropy increments $\Delta S$, that is, the exchange of heat with the surrounding medium. 

For the intracellular conditions of many plants or microorganisms, the energy release $\Delta G$ for each hydrolyzed ATP molecule  is around 50 kJ kg$^{-1}$ or about 0.5 eV\,\cite{ATP}. Therefore, the 16 ATP's in (\ref{na2}) would correspond to about 
8 eV of energy, pretty close to the 9.8 eV dissociation energy of the N$_2$ molecule. This is about half the chemical energy released by the oxidative metabolism of a glucose molecule\,\cite{ATP} and relatively costly to the energy budget of the organism. For comparison, only three ATP molecules, and two molecules of NADH (the reduced form of nicotinamide adenine dinucleotide)  are needed to incorporate a CO$_2$ molecule into  a  sugar molecule\,\cite{Calvin}. 

Nitrogenase is inactivated by oxygen, O$_2$, so nitrogen-fixing organisms have evolved various ways to exclude O$_2$  from contact with nitrogenase.
Legumes enclose symbiotic, nitrogen-fixing bacteria in root nodules from which oxygen is removed with the aid of hemoglobin molecules, similar to those that transport O$_2$ molecules in human blood.  Oceanic cynanobacteria (blue-green algae) isolate the nitrogenase in special {\it heterocysts}, thick-walled cells which have the same function as root nodules. And some photosynthesizing microorganisms only fix nitrogen after sundown when photosynthesis stops flooding the organism with O$_2$ molecules\,\cite{Bernhard}.

In water, most ammonia molecules rapidly convert to positively charged ammonium ions, NH$_4^+$, in the reversible reaction
\begin{equation}
 \hbox{NH$_3$}+ \hbox{H$_2$O}\to \hbox{NH$_4^+$}+\hbox{OH$^{-}$}.
\label{na4}
\end{equation}
As indicated by (\ref{na4}), ammonia is a strong chemical base and releases negative hydroxyl ions into solution.

In oxygenated soils or water, there are specialized microorganisms, {\it ammonia oxidizers}, that  convert ammonium ions to  nitrite ions, NO$_2^-$.  The detailed biochemistry involves several intermediate steps, but the overall reaction can be written as
\begin{equation}
\hbox{NH$_4^+$}+\frac{3}{2}\hbox{O$_2$}+2\hbox{OH$^{-}$}\to \hbox{NO$_2^-$}+3\hbox{H$_2$O}. 
\label{na6}
\end{equation}
Other microorganisms are able to oxidize the nitrites produced by reaction (\ref{na6}) to nitrate ions NO$_3^-$, as described by the overall reaction
\begin{equation}
\hbox{NO$_2^-$}+\frac{1}{2}\hbox{ O$_2$}\to\hbox{NO$_3^-$} . 
\label{na8}
\end{equation}
Both reactions (\ref{na6}) and (\ref{na8}) involve complicated biochemical pathways. Microorganisms which drive reactions (\ref{na6}) and (\ref{na8})  are able to use the modest amounts of chemical energy released to grow and multiply, although rather slowly compared to microorganisms with more energy-rich lifestyles that use O$_2$ to oxidize organic compounds to carbon dioxide and water\,\cite{Bernhard}.

\begin{figure}[h]\centering
\includegraphics[height=100mm,width=.5\columnwidth]{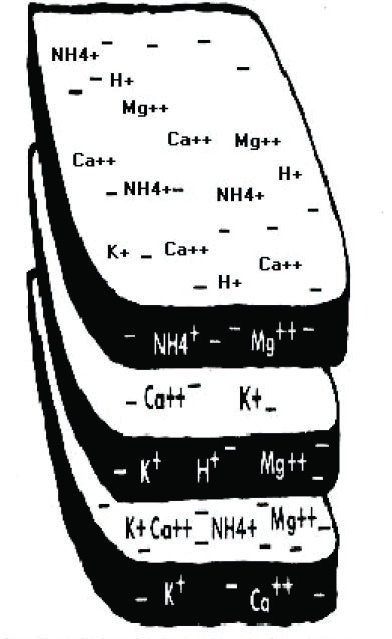}
\caption{\it The faces and edges of clay particles are covered with insoluble negative charges, to  which soluble positive cations can temporarily bind. These include many mineral nutrients of plants, including  ammonium, NH$_4^+$, potassium, K$^+$, calcium, Ca$^{++}$ and magnesium Mg$^{++}$ ions.  From reference\,\cite{clay}.
\label{cation}}
\end{figure}

The {\it nitrification} of ammonium ions to nitrate ions described by (\ref{na6}) and (\ref{na8}) is especially important in soils. Both the mineral and organic constituents of soil contain many {\it cation exchange sites}. These are sites with insoluble negative charges to which dissolved, positively charged ions or cations, like the ammonium ion, can  bind. Humus is rich in carboxylic exchange sites, -COO$^-$. Fig. \ref{cation} shows a schematic clay particulate, an aluminosilicate mineral with many insoluble negative charge sites on its faces and edges.  Cations of minerals needed by plants, potassium ions (K$^+$), calcium ions (Ca$^{++}$), magnesium ions (Mg$^{++}$) and ammonium ions (NH$_4^+$) can reversibly bind to these cation exchange sites.

There are far fewer anion exchange sites than cation exchange sites in soils, so nitrate anions can diffuse to plant roots more rapidly than ammonium ions which are held back on cation exchange sites. Within the plant tissue, nitrate ions are converted back to nitrite ions and then to ammonium ions for biosynthesis into amino acids and other nitrogen-containing organic molecules like those of Fig. \ref{N-life}.

If the nitrogen cycle only involved the reactions (\ref{na2}) -- (\ref{na8}), large and potentially toxic concentrations of ammonium, nitrite and nitrate ions would build up in soils and waters.  In fact, the buildup of inorganic nitrogen is limited by the biological process of {\it denitrification} which converts nitrogen-containing ions back into diatomic nitrogen molecules, N$_2$, and a few percent of nitrous oxide molecules, N$_2$O. These are released back into the atmosphere.   Denitrification is desirable for sewage treatment plants, or to prevent eutrophication of water bodies, or to eliminate oxides of nitrogen from the exhaust gases of high-temperature combustion in air. But denitrification is  undesirable for agriculture since it wastes the nitrogen fertility of soils. 

Denitrification is most intense in oxygen-poor soils or waters. When O$_2$ is in short supply, for example, in waterlogged soils, some microorganisms can use nitrite, NO$_2^-$,  or nitrate, NO$_3^-$, to obtain chemical energy from oxidizing organic compounds to CO$_2$, H$_2$O, N$_2$ and N$_2$O.  For a glucose molecule, C$_6$H$_{12}$O$_6$, obtained from the cellulose of plant remains, a representative denitrification process can be summarized as
\begin{equation}
8\,\hbox{NO$_2^-$}+ 8\hbox{H}^{+} +\hbox{C$_6$H$_{12}$O$_6$}\to 4\hbox{N$_2$}+6\,\hbox{CO$_2$}+10\,\hbox{H$_2$O},
\label{na12}
\end{equation}
or infrequently
\begin{equation}
7\hbox{NO$_2^-$}+\hbox{NO$_3^-$}+8\hbox{H}^{+}+\hbox{C$_6$H$_{12}$O$_6$} \to 3\hbox{N$_2$}+\hbox{N$_2$O}+6\,\hbox{CO$_2$}+10\,\hbox{H$_2$O}. 
\label{na14}
\end{equation}
Microorganisms implement oxidations, like (\ref{na12}) and (\ref{na14}), with complicated biochemical steps.

In the 1990's the importance of anaerobic ammonia oxidation (anammox) by bacteria was first recognized\,\cite{Bernhard}. In the anammox process, microorganisms use NO$_2^-$ and NO$_3^-$ ions to gain chemical energy by oxidizing ammonium ions, NH$_4^+$ instead of carbon-containing organic compounds. Representative overall reactions can be described by the equations
\begin{equation}
\hbox{NH$_4^+$}+\hbox{NO$_2^-$}\to \hbox{N$_2$}+2\,\hbox{H$_2$O},
\label{na16}
\end{equation}
or infrequently
\begin{equation}
\hbox{NH$_4^+$}+\hbox{NO$_3^-$}\to \hbox{N$_2$O}+2\,\hbox{H$_2$O}. 
\label{na18}
\end{equation}
The chemical energy released in reactions like (\ref{na16}) and (\ref{na18}) allows slow growth of the anammox microorganisms. Many of the NO$_2^-$ ions involved in the anammox oxidation process are produced by archaea, non-bacterial prokaryotes that are adapted to extreme environments\,\cite{archaea}.
Natural anammox processes are widespread  in soils and waters, and they contribute significantly to global denitrification.  Anammox systems have also been commercialized for treatment of wastewater\,\cite{anammox}.

Quantitative rates of nitrogen fixation are not known very well, but the mass of atmospheric nitrogen naturally fixed by microorganisms is believed to be approximately\,\cite{fixation}
\begin{equation}
\frac{dM_{{\rm N}_2}}{dt} \approx 175 \hbox{ Tg y}^{-1}.
\label{na19a}
\end{equation}
Here, the unit Tg y$^{-1}$ is a tera gram ($10^{12}$ g) of nitrogen per year, or a million tonnes of nitrogen per year.  Assuming that this fixation rate is very nearly the same as the denitrification rate, some 175 Tg y$^{-1}$ of nitrogen must be returned to the atmosphere by denitrification. 

Yang {\it et al.}\,\cite{Yang} estimate that natural emissions of N$_2$O are about 
\begin{equation}
\frac{dM_{{\rm N}_2{\rm O}}}{dt}=9.7\hbox{ Tg y$^{-1}$}. 
\label{na19b}
\end{equation}
Equations (\ref{na19a}) and (\ref{na19b})  imply that  denitrification produces about $(175-9.7)/9.7 \approx 17 $ N$_2$ molecules for  every one N$_2$O molecule.

It is much harder to estimate anthropogenic emissions of N$_2$O than emissions of CO$_2$, where there are reliable figures for the annual usage of fossil fuels and of cement, the main human sources. But estimates of anthropogenic N$_2$O emission rates by different authors are
\begin{equation}
\frac{dM_{{\rm N}_2{\rm O}}}{dt}=\left \{\begin{array}{ll}
7.3\hbox{ Tg y$^{-1}$},&\hbox{Yang {\it et al.}\,\cite{Yang}}\\
7.8\hbox{ Tg y$^{-1}$}  , &\hbox{Reay\,\cite{Reay} } \end{array}\right \} .
\label{na19c}
\end{equation}

The growth of nitrous oxide shown in Fig. \ref{dCdt} and in Table \ref{table4} implies that the mass $M_{\rm N}$ of nitrogen in atmospheric  N$_2$O molecules is increasing at a rate
\begin{eqnarray}
\frac{dM_{\rm N}}{dt} &=& 4\pi r^2\hat N \left(\frac{2 \,m_{\rm N}}{N_{\rm A}}\right)\frac{ dC^{\{i\}}}{dt} \nonumber\\
&=&4.1 \hbox{ Tg y}^{-1}.
\label{na20}
\end{eqnarray}
Here, $r=6371$ km is the radius of the Earth, $\hat N= 2.15 \times 10^{29}$ m$^{-2}$ is the column density of all air molecules, already given by (\ref{cm8}), $N_{\rm A}=6.02 \times 10^{23}$ is Avogadro's number, the number of molecules in a gram molecular weight, and $m_{\rm N}=14 $ g is the gram molecular weight of nitrogen atoms. The rate of increase of the N$_2$O concentration, $dC^{\{i\}}/dt= 8 \times 10^{-10}$ y$^{-1}$, taken from Fig. \ref{dCdt}, was listed in Table \ref{table4}.  

The observed growth of atmospheric N$_2$O from (\ref{na20}) is about half of the estimated human contribution to the emission rate (\ref{na19c}) of N$_2$O.  It is natural to wonder where the other half goes. Observations show that only about half of the concentration increases of atmospheric  CO$_2$ due to human emissions remains in the atmosphere after a few years. Most of the other half is thought to be absorbed in the oceans which contain about 50 times more CO$_2$ than the atmosphere. In the oceans, most molecules of the acidic oxide, CO$_2$ are reversibly converted to bicarbonate HCO$_3^-$ and carbonate ions CO$_3^{- -}$.  

Molecules of the neutral oxide N$_2$O do not form analogs of bicarbonate and carbonate ions. Therefore, in equilibrium a much smaller fraction of N$_2$O is contained in the oceans.  Weiss and Price \,\cite{Weiss} have measured the solubility coefficient $K_0$ of N$_2$O in water as a function of salinity and temperature. A representative value for the oceans is $K_0 = 4.0 \times 10^{-2}$ mole kg$^{-1}$ atm$^{-1}$. Using the known masses of the atmospheric gas and ocean water, and assuming a concentration of $C^{\{i\}}= 340 \times 10^{-9}$ molecules of N$_2$O per molecule of air, in accordance with Table \ref{table4}, this value of $K_0$ implies an equilibrium partitioning of  77\% of N$_2$O in the atmosphere and 23\% in the oceans.   For comparison, in equilibrium the partitioning for  CO$_2$  is approximately 2\% in the atmosphere and 98\% in the oceans\,\cite{pH}.  Observations\,\cite{measurements} show that the concentrations of N$_2$O in the oceans can be orders of magnitude larger than values that would be in equilibrium with the atmosphere.  Just as in the soil, biological processes in the ocean, like the anammox reactions discussed above, are creating N$_2$O, especially in suboxic waters\,\cite{measurements}.

Using the N$_2$O concentration $C^{\{i\}}= 3.4\times 10^{-7}$ of Table \ref{table4}, and other parameters the same as those for (\ref{na20}), we find that
the total mass $M$ of nitrogen contained in atmospheric N$_2$O is
\begin{eqnarray}
M_{\rm N} &=& 4\pi r^2\hat N \left(\frac{2\,m_{\rm N}}{N_{\rm A}}\right)C^{\{i\}}\nonumber\\
&=&1734 \hbox{ Tg }.
\label{na22}
\end{eqnarray}
If we assume an atmospheric residence time $\tau^{\rm N_2 O}=109$ y, in accordance with estimates by Forster {\it et al.}\,\cite{Forster}, the equilibrium loss rate of atmospheric N$_2$O would be no greater than
\begin{eqnarray}
\frac{dM_{\rm eq}}{dt} &=&M/\tau^{\rm N_2O}\nonumber\\
&=&15.9 \hbox{ Tg  y}^{-1}.
\label{na24}
\end{eqnarray}
As indicated in Fig. \ref{Ncycle}, most of the loss rate (\ref{na24}) is believed to be due to photodissociation of N$_2$O molecules by solar ultraviolet light in the stratosphere\,\cite{photodissociation}.
\section{Nitrogen in Agriculture \label{na}}
Since the biosphere is the main source of the minor greenhouse gases, nitrous oxide and methane, agriculture has been targeted with various regulations that will supposedly ``save the planet" from climate change. The planet is not in danger from greenhouse gases. But some of the regulations to address this non-problem  are of great concern since they will drastically cut food supplies for the world.

The huge increases in agricultural productivity since the year 1950 have eliminated the deadly famines that plagued mankind throughout recorded history. A number of factors have contributed to this {\it green revolution}\,\cite{Smil}, including more efficient agronomic practices,  increased concentrations of atmospheric carbon dioxide\,\cite{Taylor},  more productive plant species like hybrid corn\,\cite{Kutka}, and greatly increased use of inorganic  fertilizers\,\cite{Egli, Maharajan}. Nitrogen fertilizers are the single most important, followed by phosphorus and potassium.  Fig. \ref{yieldN} shows the strong correlation of  global cereal yield and the world production of nitrogen fertilizer.  

\begin{figure}[h]\centering
\includegraphics[height=100mm,width=1\columnwidth]{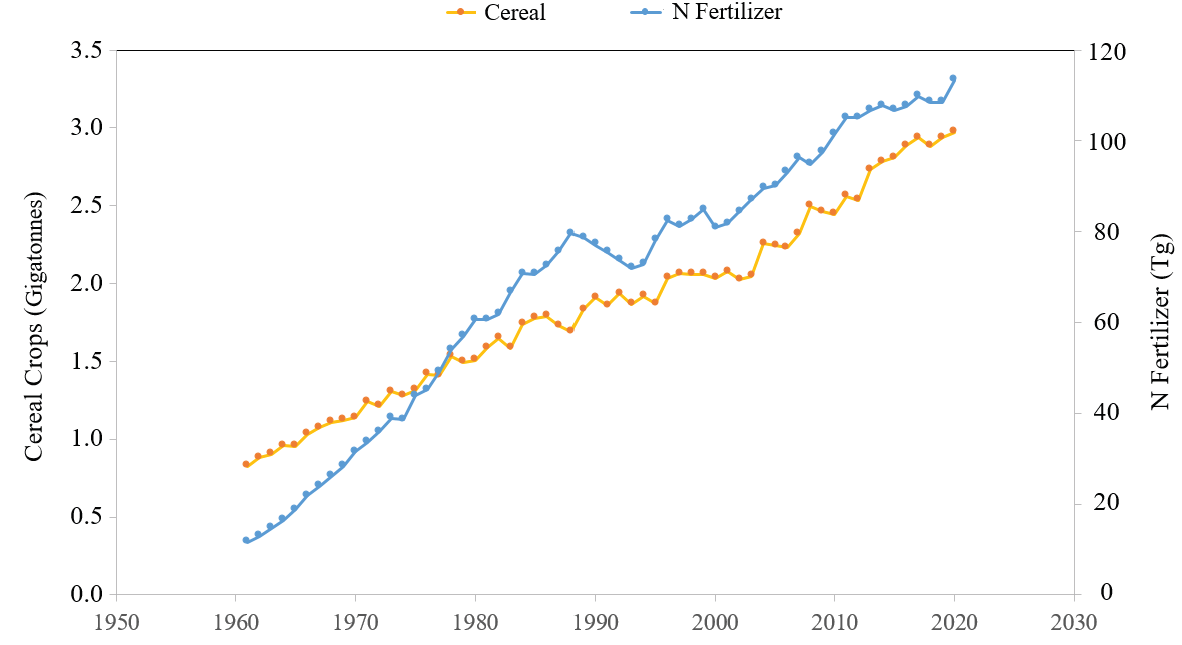}
\caption{\it Annual world production of nitrogen fertilizer used in agriculture (blue, in Tg) and world production of all cereal crops (orange, in gigatonnes) from 1961 to 2019. Data from reference \,\cite{FAOSTATS}.  The threefold increase of cereal crop yields was largely due to the use of mineral nitrogen fertilizer. Additional contributors to the increased yields were other mineral fertilizers like phosphorus and potassium, better plant varieties like hybrid corn, increasing concentrations of atmospheric CO$_2$, {\it etc.} 
\label{yieldN}}
\end{figure}

\begin{figure}[t]\centering 
\includegraphics[height=100mm,width=1\columnwidth]{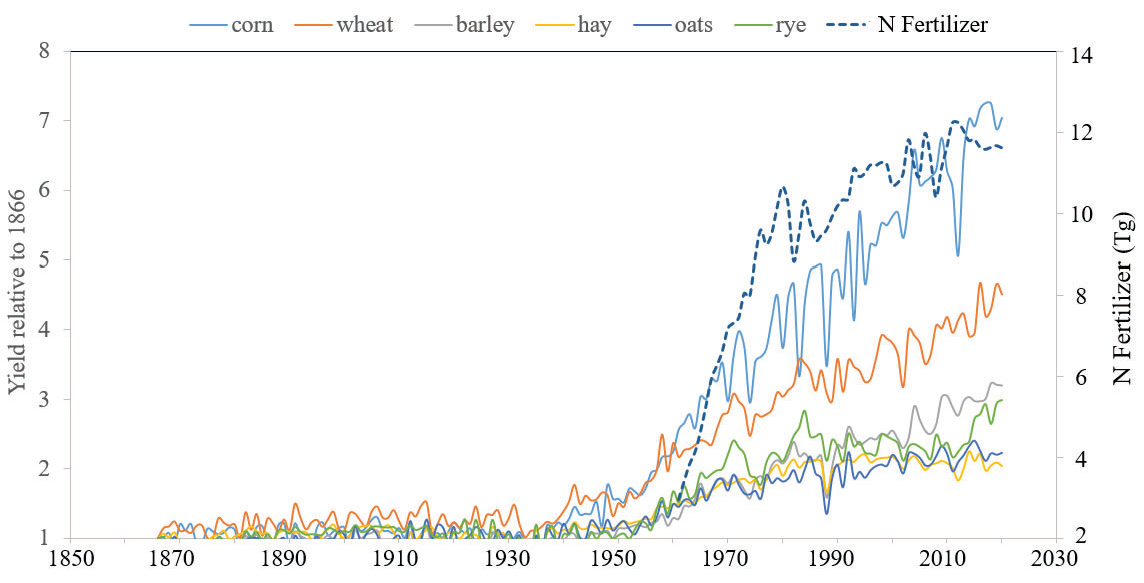}
\caption{\it Crop yields relative to yields in 1866 for corn, wheat, barley, grass hay, oats and rye in the United States. Also shown from the year 1961 is the annual mineral nitrogen fertilizer (in Tg  = megatonnes) used in agriculture. Crop yields are from the USDA, National Statistical Service\,\cite{USDA} and nitrogen fertilizer usage is from the Food Agriculture Organization statistical database\,\cite{FAOSTATS}.  Note the high correlation between yields and the use of nitrogen fertilizer.
\label{crops}}
\end{figure}

Growing plants require  many chemical elements. The largest demand is for hydrogen (H) and oxygen (O) in water, followed by carbon (C) and nitrogen (N). An optimally fertile soil must also contain phosphorus (P), potassium (K), calcium (Ca), magnesium (Mg), sulfur (S), and traces of many additional elements\,\cite{Forbes}.  

Soils naturally lose their fertility with time. Even with no removal of minerals as a result of harvested crops or logging, rains gradually wash away accessible minerals to  subsoils or to waterways. For this reason, soils in areas of high rainfall, most notably geologically old 
tropical soils, have very low fertility, often only enough for one or two years  of crops\,\cite{rainforests}.  Like phosphorus, potassium,  and other essential minerals, mineralized nitrogen can be leached from the soil by rainfall. But nitrification and the direct release of N$_2$ and N$_2$O to the atmosphere are just as important as rainfall. No analog of nitrification accelerates the loss of other minerals.

At the beginning of the 20th Century, the majority of nitrogen for crops was supplied by
manure, nitrogen fixation by soil microorganisms or legumes. Inorganic fertilizer in the form of nitrate salts or ammonia from coke ovens \,\cite{Smil} was used in small amounts. This was equivalent to only about 2\% of all nitrogen removed by crops \,\cite{Smil}. Soil nitrogen was still a significant limitation on crop production, even with the emergence of hybrid varieties. 

Commercial production of fertilizer from the reaction of nitrogen and hydrogen gas at high temperatures and pressures, in the presence of appropriate catalysts, was developed in Germany between 1910 and 1915.  Invented by Fritz Haber, for which he received the 1918 Nobel Prize in Chemistry, and commercialized by Carl Bosch\,\cite{Smil}, the Bosch-Haber process is used to produce most nitrogen fertilizer today.  Fertilizer  production did not begin in earnest until after World War II\,\cite{Smil}.  Some of the impressive consequences are shown in Fig. \ref{crops}. The yields of all non-legume food and fodder crops have increased dramatically since the year 1950, in large part due to nitrogen fertilizers.
\subsection{Efficient use of nitrogen fertilizer}
All crops need nitrogen to produce essential organic molecules like those of Fig.
\ref{N-life}. Adequate nitrogen is important throughout the growth and maturing process \,\cite{Hawkesford}. More nitrogen increases canopy development, providing opportunity for greater photosynthesis and greater crop development. Adequate nitrogen at specific times in plant development enhances growth and crop yield. Organic production, ignoring use of inorganic fertilizers, and a return to  low input agriculture, will not achieve the food supply needed to support 8.5 to 10 billion people\,\cite{Smil}. 
 
Farmers have to pay for fertilizer.  Nitrogen fertilizer is particularly expensive. Therefore, farmers have every incentive to increase
nitrogen use efficiency (NUE), the  fraction of nitrogen from fertilizer that is taken up by a crop. The lower the crop uptake of nitrogen, the greater will be the potential losses of nitrogen to air and water, and the larger the fraction of  fertilizer expenditures that are simply wasted. Typical NUE for corn production in the US is 40\% to 50\%\,\cite{Snyder}. However, with good management practices NUE can be increased to 70\%.  This represents a large financial savings and a large reduction of nitrogen emissions.

With adaptation of improved crop varieties and agronomic practices, it is difficult to separate the benefit of nitrogen fertilizer alone on yield increases. However, throughout the United States, England, Brazil and Peru, agricultural experiment stations have maintained crop plots with varying inputs of livestock manure, and inorganic N, P, K, in addition to no added fertilizer or manure amendments\,\cite{Stewart}. Stewart {\it et al.}\,\cite{Stewart} summarized plots for sites in the United States, United Kingdom and South America which compared no fertilizer versus fertilizer over long time periods. Fertilizer increased yields often by 60\% or more, compared to fields with no applied fertilizer. Stewart {\it et al.} \cite{Stewart} concluded that estimates of improved yields of 30\% to 50\% due to fertilizer use appear reasonable. 
The question becomes: What is the most economic and environmentally friendly fertilizer use strategy?

Maharjan {\it et al.}\,\cite{Maharajan} reported on the effects of incremental amounts of nitrogen fertilizer compared with no fertilizer on continuous corn plots from 1952 to 2012. Plots with no added nitrogen had yield increases of 31 kg/ha,  which represents the improvements hybrid selection and agronomic practices may achieve. Incremental amounts of nitrogen fertilizer of 45, 90, 135, and 180 kg-N/ha increased yields 1.88, 2.90, 2.95, and 1.80-fold, compared to the non-fertilized plots. Plots which received manure from beef feedlots increased yields 2.86-fold relative to the non-fertilized plots. Adding fertilizer to plots receiving manure added no benefit. However, the manure nitrogen applied was from 248 kg-N/ha to 769 kg-N/ha. These data demonstrate nitrogen fertilizer greatly increases yields.  However, supplemental nitrogen above that needed for crop growth increases the risk for environmental losses of nitrogen in leachate of NO$_3^-$, volatilization of NH$_3$ and N$_2$O, depending on soil conditions and rainfall.  Losses to the environment are costly to the ecosystem and to the farmer. Therefore, farmers are increasingly employing precision practices to improve NUE.

The ``4Rs'' describe good fertilizer management: the right source, the right amount, at the right time, and at the right place\,\cite{Snyder}. Providing needed amounts of fertilizer at optimal times for crop, soil and climatic conditions  is called precision agriculture
\,\cite{Smil, Lawrence}. For optimal NUE, adequate phosphorus and potassium are necessary\,\cite{Snyder}. Soil tests for phosphorus and potassium can determine if soil concentrations are adequate,  to ensure that response to nitrogen fertilizer will not be inhibited. The moisture content, carbon-to-nitrogen ratios in soil organic matter, pH, and soil texture, in addition to crop type, can influence the utilization of added nitrogen. Detailed farm soil maps are needed to define where soils need various amounts of added nitrogen. Accurately calculating crop nitrogen needs is a complicated process because so many variables are involved. Prudent farmers follow recommended application rates and timing, modified based on crop history and the applications of manure and mineral fertilizers in prior years\,\cite{Forbes,Connor}.

Timing of application is critical. Typical application across the US occurs in the fall, in the spring prior to planting, at planting, and after crop emergence\,\cite{Cao}. With fall application, nitrogenous losses will vary depending on cover crop. Risks of nitrogen losses to the environment are lower with application prior to planting and additional applications later in the growing period. Splitting applications into several periods can improve NUE and crop response.
\section{Conclusion}
Variants of the nitrogen cycle have been operating throughout the several billion years that life has existed on Earth;  the atmosphere has always contained some N$_2$O. There is no evidence that the concentrations of N$_2$O have ever been constant in time.  Measurements of N$_2$O concentrations in air bubbles of ice cores from Antarctica and Greenland \,\cite{N2Oice} show large changes in the concentration of N$_2$O between glacial maxima and interglacial periods like the current one.   At present, sources are emitting slightly more N$_2$O than is being absorbed by sinks. This is the reason for the upward trend of atmospheric N$_2$O that is shown in Fig.  \ref{dCdt}.  The rates of increase may look large, but all graphs, except that for refrigerant gases, have expanded vertical scales. The growth rates are relatively small, as quantified by the ``doubling times'' $t_2$ of (\ref{ff9}) that are listed in Table \ref{table4}. At current rates of increase,  it would take over 400 years to double N$_2$O concentrations. 

We use radiative forcings to predict the absolute warming of all greenhouse gases, and also the relative contributions of CO$_2$, CH$_4$ and N$_2$O.
A major conclusion of this paper is that observed rates of increase of N$_2$O pose no threat whatsoever to climate. This is based on the well-established physics of radiative transfer, which shows  that for current growth rates, the contribution of N$_2$O to warming is only 
about 6\% that of all greenhouse gases. We estimate that the absolute warming rate from N$_2$O is about 0.064 C/century.

 Few realize that large increases of greenhouse gas concentrations produce  very small increases of radiative forcing, or equivalently, small decreases of radiation to space.  Fig. \ref{N2O} shows how little radiation to space changes from the present value, the jagged black line, if N$_2$O concentrations are doubled to give the jagged red line. The black and red lines differ only slightly for frequencies near 588, 1285 and 2224 cm$^{-1}$, the normal-mode frequencies of the N$_2$O molecule shown in Table \ref{table1}.  Because forcing changes are such a small fraction of the absolute radiation to space, on the order of a percent or less for doubling greenhouse gas concentrations, the rate of temperature increase $d T^{\{\rm gg\}}/dt$ with time $t$,  must be proportional to the rate of forcing
increase $dF/dt$ with time as quantified by our Eq. (\ref{ff26a}).

Radiative forcings of greenhouse gases can be calculated very accurately on laptop computers. Elaborate general circulation models are not needed. The key question that cannot be resolved, even with supercomputers, is how much warming is needed to correct imbalance, induced by growing greenhouse gas concentrations,  between heating of the Earth by the Sun and cooling of the Earth by thermal radiation to space. This is described quantitatively by the factor $\partial T/\partial F$ of (\ref{ff26a}).  Our estimate (\ref{ff27}) of $\partial T/\partial F$, chosen to be consistent with observed temperature changes, is very close to the ``Planck value'' of (\ref{ff28}), the simplest theoretical estimate.

For present greenhouse gas concentrations, the radiative forcing per added N$_2$O molecule is about 230 times larger than the forcing per added CO$_2$ molecule.  This is due to the strong saturation of the absorption band of the relatively abundant greenhouse gas CO$_2$, compared to the much smaller saturation of the absorption bands of the trace greenhouse gas N$_2$O.  Fig. \ref{dCdt} or Table \ref{table4} show that the observed CO$_2$ rate of increase, about 2.5 ppm/year, is about 3000 times larger than the N$_2$O rate of increase, 0.0085 ppm/year.  So, the contribution of nitrous oxide to the annual increase in forcing is 230/3000 or about 1/13 that of CO$_2$.  

As discussed in Section \ref{bio}, the biosphere is the main source of atmospheric N$_2$O. Microrganisms of soils and oceans fix atmospheric nitrogen N$_2$  from the air as ammonium ions NH$_4^+$, which are subsequently converted to inorganic nitric oxide NO$_3^-$ and other compounds. These in turn are incorporated into organic molecules -- most importantly the amino acids and proteins of  living organisms. 
The nitrogen cycle, outlined in Fig. \ref{Ncycle}, describes how nitrogen from the remains of plants and other living things, from manure, mineral fertilizers, {\it etc.},  is converted many times from mineral to organic forms in soils, waters and living organisms.  Denitrification eventually returns nitrogen to the atmosphere, mostly as N$_2$ molecules, but with a small fraction of N$_2$O. 

The contributions of mineral fertilizer to the nitrogen cycle appear to be comparable to natural contributions.  Some of the slow increase of N$_2$O concentrations in the atmosphere shown in Fig. \ref{dCdt} may, therefore, be due to nitrogen fertilizer usage. How nitrogen fertilizer and natural nitrogen fixation modify the nitrogen cycle is still not completely clear and is a subject of ongoing research. But as we explained above, the details of the nitrogen cycle, though of great scientific interest, are of no concern for climate because the warming from N$_2$O is so small.  

Since few  citizens realize that the effects of total N$_2$O emissions on climate are negligible,  many governments are under pressure to ``do something" about agricultural contributions of N$_2$O.
Ideologically driven government  mandates on agriculture have usually led to disaster. The world has just witnessed the collapse of the once bountiful agricultural sector of Sri Lanka as a result of government restrictions on  mineral fertilizer\,\cite{SriLanka}. An earlier example is the collectivization of agriculture in the Soviet Union\,\cite{golodomor}, when the {\it kulak} (the derogatory Bolshevik word ``fist'' for a successful farmer) was ``eliminated as a class.'' In consequence, millions died of starvation.  Folk memories of the {\it Golodomor} (hunger-murder) played no small part in unleashing the present war in Ukraine. 

Despite this lamentable history, various governments have proposed burdensome regulations on farming, ranching and dairying  to reduce emissions of N$_2$O.  The regulations will have no perceptible effect on climate, but some of them will do great harm to agricultural productivity and world food supplies.  As we pointed out in Section \ref{na}, one of the major factors for the world's  unprecedented abundance of food in recent years has been the use of mineral nitrogen fertilizers. It is not possible to maintain highly productive agriculture without nitrogen fertilizer.  

Although agricultural emissions of N$_2$O are no threat to climate, they can lead to eutrophication of waterways. So, nitrogen fertilizer should be used intelligently to maximize nitrogen use efficiency by crops. To limit production costs, leading farmers already use no more nitrogen fertilizer than needed\,\cite{grist}.  Farmers are much better qualified to make this judgement than bureaucratic ideologues. 
 
Mandates to reduce animal numbers and fertilizer use  will  reduce agricultural yields. To continue to feed the world's growing population without mineral and natural fertilizers, agricultural areas would have to increase and encroach on native habitats, which could have remained untouched with rational use of fertilizer. The result would be more environmental stresses, not less.

\section*{Acknowledgements}
The Canadian Natural Science and Engineering Research  Council provided financial support of one of the contributing authors. This paper was first published by the CO2 Coalition, https://co2coalition.org/, in November, 2022.

\end{document}